\documentclass{IEEEtran}
\usepackage{cite}
\usepackage{amsmath,amssymb,amsfonts}
\usepackage{algorithmic}
\usepackage{graphicx}
\usepackage{textcomp}
\def\BibTeX{{\rm B\kern-.05em{\sc i\kern-.025em b}\kern-.08em
    T\kern-.1667em\lower.7ex\hbox{E}\kern-.125emX}}
\usepackage{xcolor}
\usepackage[normalem]{ulem}
\usepackage{hyperref}

\begin{document}
\title{Harnessing Selective State Space Models to Enhance Semianalytical Design of Fabrication-\\Ready Multilayered Huygens' Metasurfaces:\\Part I – Field-based Semianalytical Synthesis}
\author{Sherman W. Marcus, \IEEEmembership{Life Member, IEEE}, 
Natanel Nissan, \IEEEmembership{Student Member, IEEE},  
\\Vinay K. Killamsetty, \IEEEmembership{Member, IEEE}, 
Ravi Yadav, \IEEEmembership{Member, IEEE}, 
\\Dan Raviv, \IEEEmembership{Member, IEEE},
Raja Giryes, \IEEEmembership{Senior Member, IEEE}, 
and Ariel Epstein, \IEEEmembership{Senior Member, IEEE} 
\thanks{This work was supported by the Israel Innovation Authority through its Metamaterials consortium.}
\thanks{Sherman W. Marcus and Ariel Epstein are with the Andrew and Erna Viterbi Faculty of Electrical and Computer Engineering, Technion, Haifa, Israel (e-mail: shermanm@technion.ac.il; epsteina@technion.ac.il).}
\thanks{Natanel Nissan, Dan Raviv, and Raja Giryes are with the Faculty of Electrical and Computer Engineering, Tel-Aviv University, Tel-Aviv, Israel (e-mail: natanel.nissan@gmail.com; raviv.dan@gmail.com; raja@tauex.tau.ac.il).}
\thanks{Vinay K. Killamsetty was with the Andrew and Erna Viterbi Faculty of Electrical and Computer Engineering, Technion, Haifa, Israel. (e-mail: vinay.killamsetty@gmail.com).}
\thanks{Ravi Yadav was with the Andrew and Erna Viterbi Faculty of Electrical and Computer Engineering, Technion, Haifa, Israel. He is now with Reykjavik University, Iceland (e-mail: raviyadav2100@gmail.com).}
\thanks{The code developed and utilized in Part I and Part II of this two-part compilation is publicly available at https://github.com/nati223/metamamba.  }}

\maketitle

\begin{abstract}
Planar metasurfaces can profoundly control electromagnetic scattering. At microwave frequencies, such devices are typically implemented using multilayer cascades of patterned metallic sheets, whose design often requires time-consuming full-wave optimization. Here, we extend analytical models originally developed for sparse loaded‑wire metagratings to accurately describe densely packed Jerusalem‑cross meta‑atoms embedded in standard printed circuit board (PCB) dielectric stacks. The model captures both near‑ and far‑field coupling within and between layers, enabling efficient prediction of the dual‑polarized response. Using this framework, we identify highly transmissive meta‑atoms whose phase is controlled by the leg lengths of the Jerusalem crosses (microscopic design stage). This (phase)-(leg-length) ”lookup table” allows rapid synthesis of Huygens’ metasurfaces (macroscopic design stage), demonstrated through a full‑wave‑validated metalens exhibiting low‑reflection beam manipulation. Notably, we implement a judicious scaling method to further extend the model to predict wideband meta‑atom responses. In the companion paper (Part II), a hybrid machine‑learning approach leverages this semianalytical framework to enhance accuracy without requiring the conventional exhaustive full‑wave training, enabling ultrafast inverse design across the full parameter space. Overall, the presented methodology -- the standalone semianlytical scheme (Part I) and the machine-learning enhanced version (Part II) -- establishes an effective open‑source toolkit for versatile, rapid, and highly accurate synthesis of fabrication-ready dual-polarized transmissive Huygens’ meta-atoms and metasurfaces.
\end{abstract}


\section{Introduction}
\label{sec:introduction}
Over the past couple of decades, metasurfaces (MSs) have flourished as an emerging alternative to traditional (passive or adaptive) bulky optical components and microwave beamforming devices \cite{glybovski2016metasurfaces}. Like their 3D metamaterial predecessors, they utilize closely packed subwavelength polarizable particles (meta-atoms) to attribute advanced and exotic macroscopic field manipulation functionalities to the devised composites. One of the key reasons for their greater popularity relative to conventional metamaterials -- to which MSs owe their growing dominance in recent years -- is their comprehensive control of electromagnetic wavefronts, and their ability to perform versatile wavefront transformations with subwavelength surface thicknesses \cite{glybovski2016metasurfaces}. 

One of the earliest classes of MSs that is also widely used to this day is the phase gradient MS. Devices belonging to this class feature meta-atoms that are designed to have unitary transmission magnitude, and transmission phase that -- when properly arranged along the surface plane -- would form a (spatially-varying) phase profile as to locally compensate for the phase difference between an incident wave and a desired scattered wave \cite{LalanneOptLett1998,LalanneOptSocAmA1999,bomzon2001pancharatnam,bomzon2002radially,HasmanAPL2003}. They are capable of performing various functionalities like anomalous reflection and refraction, polarization conversion, and focusing, among many others \cite{YuScience2011,aieta2012out,lin2014dielectric,Yu2014}. 
%

Addressing MS analysis and synthesis rigorously via the generalized sheet transition conditions (GSTCs) \cite{tretyakov2003analytical,Kuester2003} -- formally relating the fields on both sides of the surface to the homogenized surface parameters -- it has become clear in recent years that such phase gradient MSs are actually a sub-class of the more general Huygens' MS (HMS) cateogry \cite{epstein2016huygens}. HMSs feature collocated electric and magnetic responses, and are correspondingly characterized by the electric surface impedance $Z_{se}$ and magnetic surface admittance $Y_{sm}$ distributions as constituent parameters; phase gradient MS response can then be induced by properly balancing these electric and magnetic polarizabilities to suppress reflections for normally incident waves (Huygens' condition \cite{LoveRadSci1976,JinIEEEAWPL2010}), while the interaction strength serves as an additional degree of freedom to induce any prescribed phase shift at the meta-atom level \cite{pfeiffer2013metamaterial,MonticonePRL2013,SelvanayagamOptexp2013,KivsharAdvOptMat2015}. While early demonstrations of corresponding reflectionless beam manipulating HMSs followed the (perhaps most natural) manifestation of such electric and magnetic responses as collocated wire and loop meta-atoms \cite{pfeiffer2013metamaterial, SelvanayagamOptexp2013,wong2014design}, it was later shown that the required surface properties of such Huygens meta-atoms can be produced by a symmmetric cascade of impedance sheets \cite{pfeiffer2013millimeter,MonticonePRL2013}. These have the great benefit of being fully compatible -- implementation-wise -- with conventional via-less multilayered printed circuit board (PCB) configurations, readily translatable to practical devices at microwave frequencies via standard fabrication techniques \cite{Pfeiffer2013Cascaded, Jiang2017Metamaterial, Kasahara2018, chenIEEETAP2019, Liu20Apertures, Chen2023Microwave, Liu2025}.

To realize such HMSs, the standard MS design procedure is typically followed, composed of two main steps. First, in the \textit{macroscopic} design stage \cite{epstein2016huygens}, the necessary surface constituent distribution along the HMS is retrieved by considering the field discontinuity required to realize the transformation of interest in conjunction with the GSTCs. The resultant spatially-varying MS specifications can be cast in the form of the aforementioned $Z_{se}(\vec{\rho})$ and $Y_{sm}(\vec{\rho})$, where $\vec{\rho}$ is a point on the HMS \cite{pfeiffer2013metamaterial,SelvanayagamOptexp2013,epstein2016cavity}; or, equivalently, as local reflection and/or transmission coefficients \cite{MonticonePRL2013, Pfeiffer2013Cascaded, AsadchyPRL2015, EstakhriPhysRevX2016,chenIEEETAP2019,Liu20Apertures}, e.g. the transmission magnitude $|T(\vec{\rho})|$ and phase $\angle T(\vec{\rho})$ as common for phase-gradient MSs. Subsequently, in the \textit{microscopic} design stage, suitable meta-atom geometries (subwavelength patterns of metallic and dielectric substances) are devised as to locally realize these prescribed abstract surface parameters at each point $\vec{\rho}$ along the HMS. This correspondence between a given configuration and its effective constituents is commonly obtained by probing the meta-atom response to plane-wave illumination when embedded in a uniform homogeneous array of identical meta-atoms, associating the recorded scattering parameters with the local meta-atom response \cite{Pfeiffer2013Cascaded,epstein2016huygens,PfeifferPhysRevAppl2014}.

While the considerations involving HMS macroscopic design concepts and methodologies are quite established by now \cite{pfeiffer2013metamaterial,MonticonePRL2013,SelvanayagamOptexp2013, WongIEEEAWPL2015, epstein2016huygens, epstein2016cavity, EpsteinPRL2016, AsadchyPRB2016, chen2018theory, lavigne2018susceptibility, ataloglou2021arbitraryy}, substantial obstacles still impede rapid execution of the microscopic design task, which is essential to produce physical PCB layouts for practical field-operative devices and applications. One such obstacle is the fact that designs often rely on time-consuming full-wave optimizations \cite{epstein2016cavity,  lavigne2018susceptibility}. Indeed, using transmission line models (TLM) to analytically predict scattering of impedance sheet cascades (corresponding eventually to multilayer PCB MSs) have been proposed in the past years \cite{MonticonePRL2013,pfeiffer2013millimeter,PfeifferPhysRevAppl2014,WongIEEEAWPL2015,EpsteinIEEETAP2016}. While such approaches dramatically reduce the required design time (and often provide invaluable physical insight), they suffer from limited accuracy stemming from neglecting near-field interlayer coupling \cite{PfeifferPhysRevAppl2014,chen2018theory,cole2018refraction}. This becomes more significant as the electrical thickness is reduced in size relative to the meta-atom lateral dimensions, as often occurs for ultracompact MSs. Such inaccuracies may sometimes be lessened by limiting the unit cell to deeply subwavelength dimensions, or by employing a laminate with a very large dielectric constant \cite{chenIEEETAP2019,AbdoTAP2019}, but these considerably restrict the design and implementation possibilities. An alternative for overcoming this interlayer coupling issue involves introducing more elaborate circuit models into the TLM to account for mutual coupling between the layers \cite{xu2018technique,olk2019accurate}.  These components, however, must be updated iteratively, or satisfy certain assumptions on impedance sheet response, thereby reducing the generality of these more accurate methods and limiting their applicability. Thus, in practice, optimization in full-wave solvers is often unavoidable for proper meta-atom design \cite{PfeifferPhysRevAppl2014,chen2018theory,cole2018refraction}. 

In this paper, we address these gaps, presenting a semianalytical scheme which eliminates the need for full-wave optimization in the microscopic stage of the design. The scheme is based on a rigorous analytical model -- codified in a Matlab program we developed called LAYERS -- for determining Floquet scattering from periodic multiple layers of dielectric substrates with etched metallic patterns, accounting for both evanescent and propagating modes of propagation, realistic conductor and dielectric loss, and considering all mutual interactions between the scatterers. The formulation is an extension of an analytical model used previously for addressing scattering from sparse loaded wire metagratings, in which the utilized unidirectional conducting strips interacted almost exclusively with the transverse electric (TE) polarization \cite{EpsteinPhysRevAppl2017,RabinovichIEEETAP2018,RabinovichIEEETAP2020}. Since the main task underlying the microscopic design stage for PCB HMSs involves associating such a physical meta-atom configuration with its transmission coefficient when embedded in a uniform periodic array and illuminated by a normally incident plane wave, properly adapted versions of the model to accommodate closely packed unit cells in a subwavelength periodic lattice, would form an effective theoretical framework to this end. While we have previously presented preliminary versions of this idea in conferences \cite{Levy2019,Killamsetty2021Conf} and applied it to single-polarized HMS designs \cite{Kuznetsov2024,MateosRuiz2025}, this is the first time that the entire method is derived and described in detail. 

Importantly, we develop and present herein two critical extensions: enabling synthesis of dual-polarized devices, and frequency response analysis, both essential for maturing the proposed methodology towards practical use in real-life applications. The former is achieved by replacing the single-polarized capacitively loaded wires with dual polarized Jerusalem cross (JC) patterns, properly adapting the semianalytical scheme to harness the leg-lengths of JCs in the various laminates as degrees of freedom for tuning the overall multilayer meta-atom response. The latter is facilitated by a judicious scaling approach which controls the size of all length parameters relative to wavelength, including the JC leg-length from which the effective wire impedance can be gleaned, providing reliable prediction of the meta-atom transmission efficiency and phase shift not only at the nominal frequency, but also over an entire frequency band. 

Considering these two extensions, the semianalytical methodology presented herein enables swift production of lookup table (LUT) entries for a given multilayer PCB stack at a given operation frequency, matching each target phase shift in the interval $-\pi$ to $\pi$ with a realistic layout of JCs defined on the prescribed laminate cascade. Validation using full-wave solvers of the individual meta-atom designs indicates good agreement between analytical predictions and simulated performance. Subsequently, a dual-polarized HMS metalens is composed as a representative case study, utilizing the devised LUT (microscopic design) and following the typical phase gradient approach (macroscopic design), revealing notable efficiency and very good focus properties in simulation.

These results establish the efficacy and appealing features of the developed scheme for \textit{on demand} design of versatile practical HMSs. Nonetheless, experience with the implemented algorithms indicates that, for some meta-atoms of the LUT, when attempting to meet certain transmission phase-shift values (requiring highly resonant meta-atom configurations), or when the interlayer spacings (laminate thicknesses) within the user-defined PCB stack are very small, the approximations underlying the analytical model might introduce discrepancies between the theoretical predictions of the effective meta-atom scattering parameters and full-wave ("ground truth") results for these parameters. In this Part I of our two-part compilation, we overcome these occasional discrepancies in the scattering parameters of the LUT meta-atoms by simply using the full-wave (CST) results for each of these 150-or-fewer meta-atoms in the LUT, a task that on a standard personal computer (PC) only requires a few minutes per meta-atom. This overall procedure -- utilizing the semianalytical model to find the meta-atom properties that provide high transmittance covering the entire $2\pi$ phase shift interval, while determining the final transmittance with the aid of CST -- is sufficient for enabling effective macroscopic design of HMSs. However,  targeting a specific phase-shift value or reaching the absolute optimal transmittance (or, for that matter, a prescribed transmittance value) cannot be guaranteed \textit{a priori}.

Consequently, we propose in Part II \cite{Nissan2026} of this two-part compilation a complementary machine learning technique (MetaMamba) to further enhance the accuracy of the semianalytical scheme, thereby reducing the need for \textit{a posteriori} full-wave validation. In particular, a unique hybrid approach is adopted, where intensive training is effectively performed using our rapid semianalytical algorithm (Part I), while a limited number of runs in a full-wave solver are used to "fine-tune" the selective state space model in order to reach near-perfect agreement with the full-wave ground truth for the entire parametric space. Thus, owing to the reliable analytical model, time-consuming full-wave simulations are kept to a minimum (in contrast to typical training procedures of common neural networks used for electromagnetic MS design \cite{Naseri2021TAP,Oliver2022reflect,Niu2023OJAP,Wang2023GAN,Liu2024_gan_fullspace}), without compromising on precision. Furthermore, once thus finalized, the resultant MetaMamba offers additional substantial computational benefits, enabling ultrafast inverse design of such PCB-compatible multilayer meta-atoms for any desired transmission phase and augmented range of transmittance values. Hence, this entire two-part series-- the standalone semianlytical scheme (Part I) and the machine-learning enhanced version (Part II) -- establishes an effective open-source \cite{github} set of previously inaccessible engineering tools for versatile, rapid, and highly accurate synthesis of transmissive Huygens’ meta-atoms and MSs.

\section{Theory}
\label{sec:theory}

\begin{figure}[!t]
\centerline{\includegraphics[width=0.8\columnwidth]{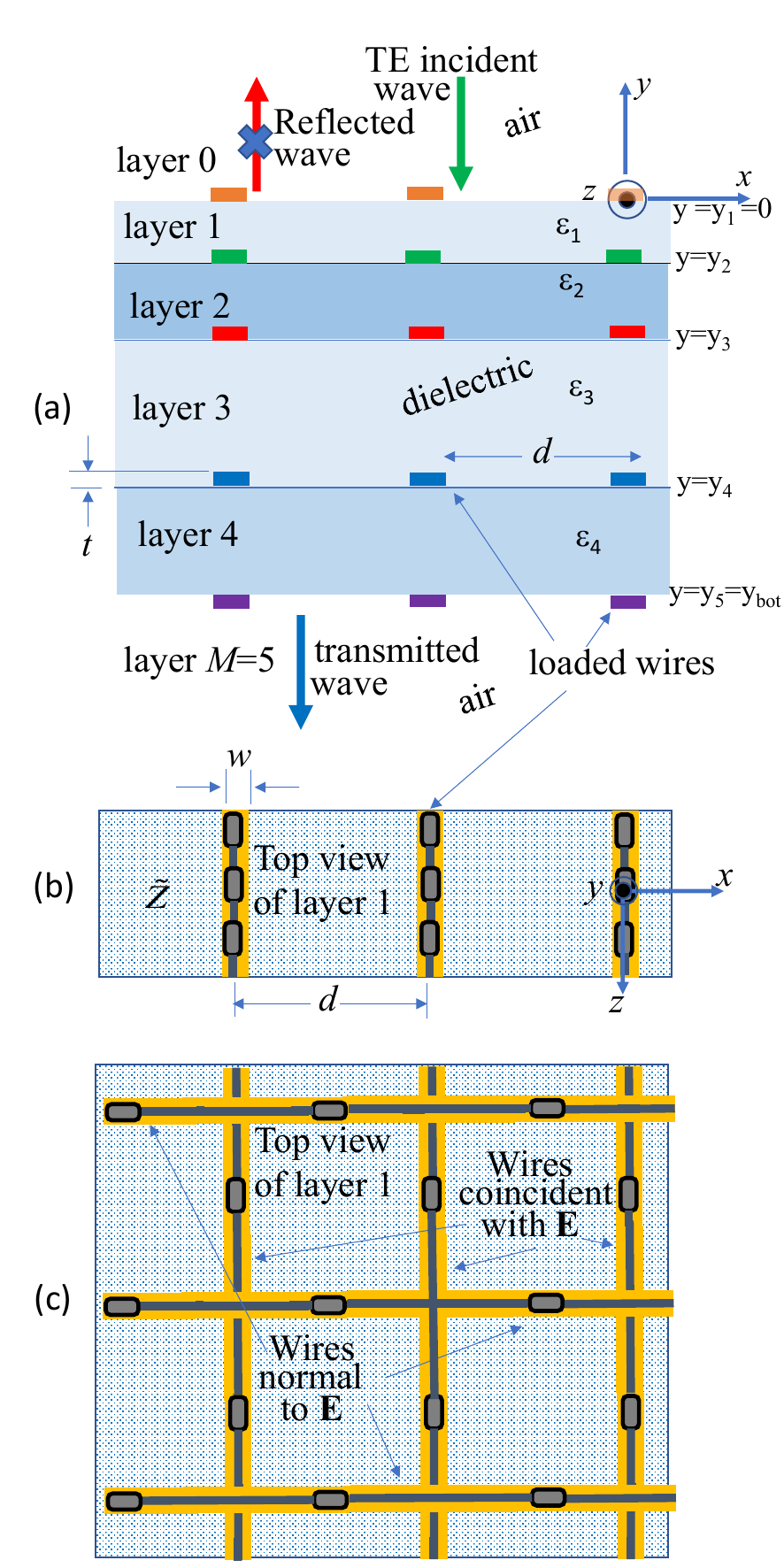}}
\caption{(a) Parallel arrays of periodic capacitively loaded wires affixed to laminates in a multilayer PCB configuration, illuminated by a transverse electric (TE) polarized plane wave. The goal to eliminate the reflected propagating wave is indicated by the cross on the corresponding red arrow. (b) The top view of layer 1, displaying the distributed impedance $\tilde{Z}$ of its wires. Although illustrated as short discrete reactive components, the wire is modeled as having a continuous, distributed impedance per unit length along the wire. For TE polarization, these wires are oriented in the direction of the electric field vector.  (c) The same surface as (b), but including additional wires resulting in $90^\circ$ rotational symmetry of the system about the $y$-axis. }
\label{figIntro}
\end{figure}
%

\subsection{Problem Formulation}
\label{sec:LayeredMediaConfig}
In order to form the desired LUT as part of the microscopic design stage mentioned in Section \ref{sec:introduction}, the transmission and reflection coefficients associated with  sub-wavelength meta-atom geometries should be evaluated, when each particular meta-atom is part of an array of identical meta-atoms. This requires a precise description of the meta-atom's configuration.

Consider the multilayer PCB configuration shown in Fig. \ref{figIntro}a, the topmost surface of which is coincident with the ${x\mathrm{-}z}$ plane. Loaded copper wires are etched periodically onto the top surface of each dielectric laminate, as well as onto the bottom surface of the lowest laminate. These wires have a width $w$ that is small relative to the wavelength $\lambda$, and a smaller thickness $t\ll w$, and their centers are laterally located at $x=0$, $x=\pm d$, $x=\pm 2d$, ... , where $d$ is the period (Fig. \ref{figIntro}b). As shown in the figure, the wire arrays are located at the interfaces between the laminates, where interface $n$ is at ${y=y_n}$. The loading is characterized by a distributed impedance (impedance per unit length), implemented by placing reactive components along the wires at distances (very) small relative to $\lambda$. Herein, the distributed impedance $\tilde{Z}_n$ along interface $n$ will often be expressed using the non-dimensional distributed impedance $Z_n$, where $\tilde{Z}_n=Z_n(\eta/\lambda)$ and $\eta$ is the impedance of free space. The dielectric constant of the $n$th laminate is $\varepsilon_n$ which is generally complex.

The layered configuration is illuminated by a plane wave normally incident from above. For TE polarization, the E-field is $z$-directed, so that the incident wave is given by
 \begin{equation}
E_{z}^{\mathrm{inc}}(y)\!\!=\!\!E_0 e^{-iky},
\label{eqEinc}
\end{equation}
where an an $e^{-i\omega t}$ time dependence is suppressed. Since the configuration is periodic, the field is composed of a discrete spectrum of plane waves governed by Floquet-Bloch (FB) theory \cite{tretyakov2003analytical}. The analytical model which determines these fields (described in the Appendix, following \cite{RabinovichIEEETAP2020}) considers full coupling of the propagating and evanescent waves in all layers of the meta-atom, as well as losses in each of the dielectric laminates\footnote{The wire strip thickness and (finite) conductivity are accounted for implicitly through the distributed impedance of the wires (see Section \ref{sec:DetZkWk}).}. 
The waves scattered in the reflected and transmitted directions are: 
 \begin{equation}
E_z^{\mathrm{ref}}(x,y)=\sum\limits_{p=-\infty }^{\infty }e^{i2\pi px/d}B_p^{(0)}e^{ik_{yp}y},y>0,
\label{eqAppend1Ref}
\end{equation}
 \begin{equation}
E_z^{\mathrm{tran}}(x,y)=\sum\limits_{p=-\infty }^{\infty }e^{i2\pi px/d}A_p^{(M)}e^{-ik_{yp}(y-y_{\mathrm{bot}})},y<y_{\mathrm{bot}},
\label{eqAppend1Tran}
\end{equation}
where the sums are over the integer FB mode number $p$, $k_{yp}=(2\pi/d)\sqrt{(d/\lambda)^2-p^2}$ is the transverse wavenumber of the $p$th mode, and the $A_p^{(m)}$ and $B_p^{(m)}$ belong to the set of unknown coefficients which describe the discrete spectrum of $E$ and $H$ fields in each layer $m$ of the meta-atom.  In particular, referring to $A_p^{(M)}$ and $B_p^{(0)}$ in \eqref{eqAppend1Ref} and \eqref{eqAppend1Tran}, layer $m=0$ is the uppermost region $y>0$ above the structure, and layer $M$ is the bottommost region $y<y_{\mathrm{bot}}$ below the structure (see Fig. \ref{figIntro}a). By imposing the required boundary conditions on these fields  -- considering transitions between dielectric layers as well as load-dependent currents induced on the wires and the associated discontinuity of the tangential field components [\eqref{eqAppend7}, \eqref{eqAppend8}, \eqref{eqAppend13}] -- we may solve for these unknown coefficients (and thus the entire scattered field solution) via the resultant system of simultaneous linear equations (see Appendix for the detailed derivation).

Some of the plane waves in the sums of \eqref{eqAppend1Ref} and \eqref{eqAppend1Tran} would be propagating and the remainder would be evanescent in accordance with whether $k_{yp}$ is real. The number of propagating waves would depend on $d/\lambda$: the greater this ratio, the greater would be the number of FB waves that are propagating. Herein, in order to be compatible with the homogenization approximation underlying conventional MS design, we choose $d/\lambda$ sufficiently small to allow only a single ($p=0$) wave to propagate in the reflected direction characterized by the reflection coefficient $R$ (specular reflection), and a single ($p=0$) wave to propagate in the transmitted direction characterized by the transmission coefficient $T$ (direct-ray transmission). From \eqref{eqAppend1Ref} and \eqref{eqAppend1Tran} it is clear that 
 \begin{equation}
T=A_0^{(M)}/E_0,\,\,\,\,\,\,\,\, R=B_0^{(0)}/E_0. 
\label{eqTRdef}
\end{equation}
Both $R$ and $T$ are complex; the upward red arrow in Fig. \ref{figIntro}a represents $|R| \le 1$, and the downward blue arrow represents $|T| \le 1$. $T$ will be of principal interest in what follows. Since the field solutions stem from a self-consistent stipulation considering the currents induced on the various loaded wires via Ohm's law [\eqref{eqAppend12}, \eqref{eqAppend13}], the correspondingly resolved 
$T$ will depend on the impedances of the wires in all the layers: 
 \begin{equation}
T=T(Z_1,Z_2,...,Z_N).
\label{eqTR}
\end{equation}
Determination of \eqref{eqTR} for a given meta-atom configuration is the main purpose of the analytical model described in the Appendix. This analytical model has been codified for Matlab as LAYERS, facilitating the compilation of a suitable LUT for the target microscopic HMS design (Sections \ref{subsec:LUT}, \ref{sec:buildingLookupTable}).
  
\subsection{Dual Polarization}
\label{sec:DualPol}

To this point, we have been considering multi-layered meta-atoms consisting of simple loaded wires (Fig. \ref{figIntro}b), compatible with the standard metagrating analysis and synthesis formalism \cite{RabinovichIEEETAP2020}. We will now expand the discussion to a more intricate form of conductor trace geometry, attributing the meta-atom with augmented field manipulation features.

Since, as stated, the E-field is directed in the $z$-direction, it is in the same direction as the orientation of the wires in Figs. \ref{figIntro}a and \ref{figIntro}b. This electric field will induce currents in these $z$-oriented wires which will in turn produce secondary $z$-directed electric fields \cite{tretyakov2003analytical}. Conversely, if additional wires are present that are normal both to the existing wires in Figs. \ref{figIntro}a and \ref{figIntro}b and to the direction of the E-field, then these additional wires (Fig. \ref{figIntro}c) will not interact significantly with the rest of the meta-atom. We may therefore neglect such cross-polarization interactions (i.e., ignore these $x$-directed wires) in the mathematical model. However, if these additional wires are arranged periodically with the same period $d$ as the original wires, and at each interface they have the same $Z_n$ as the original wires, then the entire system possesses $90^\circ$ rotational symmetry. As such, if the entire system is rotated $90^\circ$, or alternatively the polarization is changed from TE to transverse magnetic (TM), the fields scattered would be identical to those scattered in the original case.  It is this "dual-polarized" configuration that will be considered herein.

\begin{figure}[!t]
\centerline{\includegraphics[width=0.8\columnwidth]{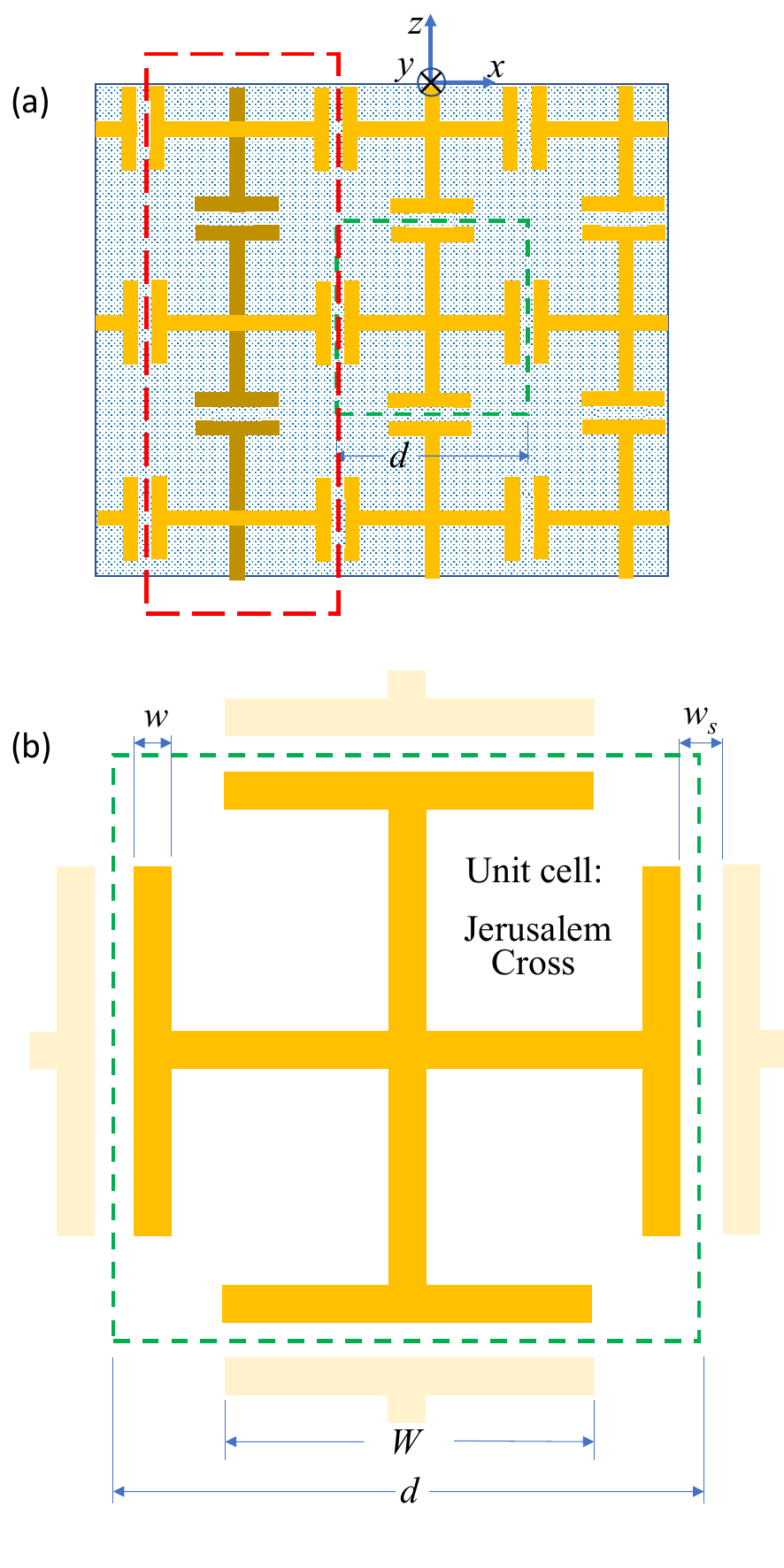}}
  \caption{(a) The reactive loading of the wires in Fig. \ref{figIntro}c represented by planar parallel plate capacitors. The unit cell is defined about a JC-shaped structure. A new "effective" loaded wire is shown in the red dashed box. The brown wires indicate the realization of the original capacitive loading in Fig \ref{figIntro}a. (b) Enlarged view of the unit cell. The distributed impedance of the "effective" wire is a function of the length $W$ of the leg of the JC.
  }
\label{figCapacitors}
\end{figure}

Although up to now we allowed any type of loading on the wires in Figs. \ref{figIntro}b and \ref{figIntro}c, we will now limit consideration to capacitive\footnote{While the formalism presented herein can be applied to any type of load (be it inductive, capacitive, resistive, or a combination of the three \cite{EpsteinPhysRevAppl2017,PopovPRAppl2019, XuTAP2021, yashno_large-period_2022, boust_metagrating_2022, tan_design_2023, Kuznetsov2024}), the highly inductive nature of the thin $w\ll\lambda$ conducting strips typically requires capacitive loading in order to drive the entire loaded-wire configuration closer to resonance, thus enabling sufficient dynamic range for tuning its response via the load degrees of freedom (capacitor width $W$ in our case, \textit{cf.} Fig. \ref{figCapacitors}b).} loading. This loading is realized by planar (printed) parallel plate capacitors identical to the brown colored wires in Fig. \ref{figCapacitors}a.  In that figure, such capacitors have been placed in each vertical wire in the spatial regions between the horizontal wires, and similarly in each horizontal wire in the regions between the vertical wires. 

\begin{figure}[!t]
\centerline{\includegraphics[width=0.6\columnwidth]{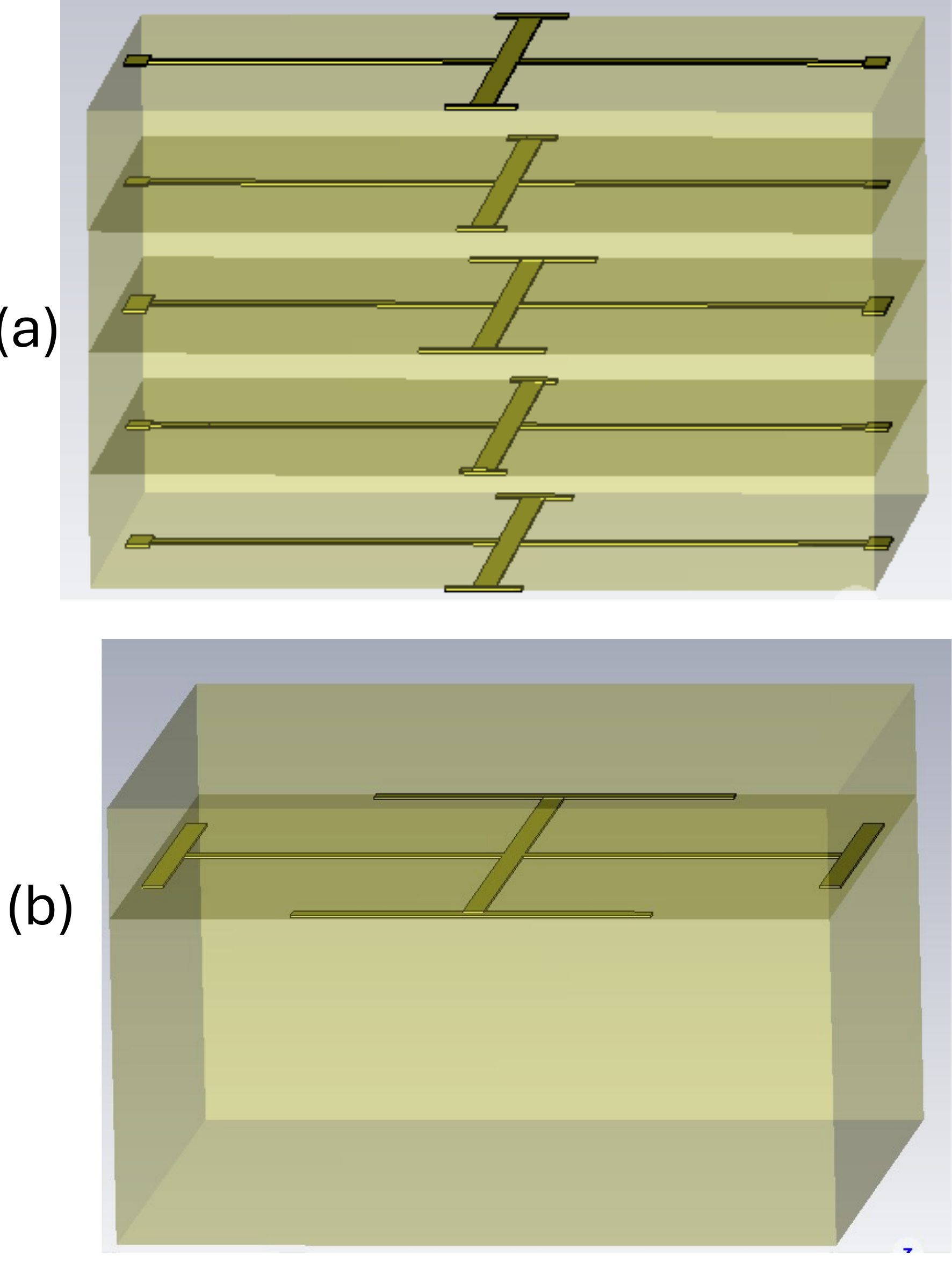}}
  \caption{(a) A multilayered unit cell containing JCs at each of the five interfaces.  The length $W_n$ of the leg of each JC can be varied to control the value of $Z_n$, and subsequently $T$ and $R$. (b) Example of a single interface ($n=2$) utilized to determine  $Z_n(W_n)$.
  }
\label{figfiveinterfaces}
\end{figure}

Although the unit cell of the resulting two dimensional array can be defined as any $d \times d$ area, it is convenient to define it in a manner in which the boundary of the unit cell does not intersect the metal. Such a unit cell is shown enclosed by the green dashed box in Figs. \ref{figCapacitors}a and \ref{figCapacitors}b. As shown in Fig. \ref{figCapacitors}b, the metal within each such unit cell has the shape of a JC\footnote{Although "Jerusalem cross" refers to a shape that is a bit more intricate than that pictured in Fig. \ref{figCapacitors}b (see https://en.wikipedia.org/wiki/Jerusalem\_cross), it has become customary to utilize the term for this shape as well \cite{padooru_analytical_2012,Armghan2025}.}, which is geometrically characterized by the wire width $w$, thickness $t$, period $d$, leg length $W$ and separation distance $w_s$ between the plates of adjacent JCs. To simplify the synthesis procedure (Sections \ref{subsec:LUT} and \ref{subsec:macroscopic_design}), we set all these parameters to be the same for all of the $N$ interfaces of the PCB stack in Fig. \ref{figIntro}a -- except for the leg-length parameter $W=W_n$, which would serve as our design degree of freedom to control the effective load impedance $Z_n$ at the $n$th layer, thus facilitating engineering of the overall multilayer meta-atom response. The unit cell of such a stack with five interfaces is illustrated in Fig. \ref{figfiveinterfaces}a.


\subsection{Effective Load Impedance}
\label{sec:DetZkWk}

In order to physically implement \eqref{eqTR}, it is necessary to relate the distributed impedance $Z_n$ to an actual wire geometry. This was accomplished for the simpler case of parallel wire strips (Figs. \ref{figIntro}a and \ref{figIntro}b) \cite{EpsteinPhysRevAppl2017} by relating $Z_n$ to the "plate" size (brown-colored wires of Fig. \ref{figCapacitors}a) of an equivalent planar parallel plate capacitor \cite{garg2024microstrip}. This "simpler" relationship between wire geometry and $Z_n$ is not advantageous for the JC geometry at hand, because this geometry features numerous capacitive channels that may be difficult to account for explicitly. Instead, a more general semianalytical method, adapted from \cite{PopovPRAppl2019}, will be utilized, relying on the observation that, despite the apparently intricate JC features, the dominant scattering from the subwavelength meta-atoms still stems from the co-polarized dipole moment developing on the wires when excited by the incident wave. In other words, we may approximate the scattered fields associated with a reference JC chain along the polarization axis as originating from an equivalent line source, the induced current on which is, again, determined by an effective uniformly distributed load impedance $Z_n$. Correspondingly, as shall be detailed below, we may deduce the effective $Z_n$ corresponding to a given JC geometry by considering the scattering coefficients off the relevant JC chain in a full-wave solver and comparing them to the ones expected for the abstract uniformly loaded configuration using the formalism presented in Section \ref{sec:LayeredMediaConfig} and the Appendix. 

Following this rationale, our specific goal in this subsection is to relate the $Z_n$ to the value $W_n$ of the JC leg-length for interface $n$, i.e., to determine the functional form
\begin{equation}
Z_n=Z_n(W_n),1 \le n \le N.
\label{eqZkofWk}
\end{equation}
Since the quasistatic ($d\ll\lambda$) distributed impedance $Z_n$ of a wire at one interface can be expected to be only negligibly dependent on the $Z_n$ at another interface, each interface will be considered separately\footnote{The restriction $d\ll\lambda$ implies interaction only through the dipole moment and neglecting higher-order multipoles. This modeling approach still provides significantly higher accuracy than the standard TLM approach that ignores any evanescent wave phenomena (\textit{cf.} Section \ref{sec:introduction}), since near-field interlayer coupling is fully taken into account in the frame of the considered dipole approximation.}. Nevertheless, the $Z_n$ at one interface \emph{can} be expected to be influenced by the wire's immediate surroundings, in particular the dielectric material in which it is immersed \cite{PfeifferPhysRevAppl2014}. Therefore, for this effective load impedance $Z_n(W_n)$ evaluation process, each interface with a JC unit cell will be considered independently, but together with all layers of the dielectric. This is illustrated in Fig. \ref{figfiveinterfaces}b for the interface located at $y=y_2$.

For each interface, the unit cell of the periodic structure is defined in a full-wave solver. In this study, CST was employed for this purpose\cite{CST}. Such solvers only require the unit cell to be defined, and the S-parameters -- corresponding to terms in the FB series representing the fields in reflection and transmission regions \eqref{eqTRdef} -- are provided as output. Correspondingly, the reflection and transmission S-parameters associated with the fundamental (zeroth-order) mode are equivalent to $R$ (specular reflection) and $T$ (direct transmission) of \eqref{eqTRdef}, respectively. For the configuration similar to Fig. \ref{figfiveinterfaces}b in which only interface $n$ is considered, the reflection and transmission coefficients will be denoted $R^{(n)}$ and $T^{(n)}$, respectively. Performing a CST sweep over incremental values of $W_n$ from $W_{min}=0$ to $W_{max}$ (determined in accordance with the unit cell dimensions), it is possible to tabulate $T_{\text{CST}}^{(n)}(W_n)$ and $R_{\text{CST}}^{(n)}(W_n)$\, where the subscript was added to indicate the computational source of the results. 

Since we may utilize LAYERS (see Appendix) to analyze any periodic stack of the form of Fig. \ref{figfiveinterfaces}a [see \eqref{eqTR}], we can of course use it to calculate $T^{(n)}(Z_n)$ for the simpler configuration including only a single layer at a time (e.g., as in Fig. \ref{figfiveinterfaces}b). Subsequently, the non-dimensional distributed impedance $Z_n$ may be obtained for each value of $W_n$ by solving
 \begin{equation}
T^{(n)}(Z_n)-T_{\text{CST}}^{(n)}(W_n)=0
\label{eqTZW}
\end{equation}
for $Z_n$, which can easily be accomplished in Matlab\footnote{As detailed in Section \ref{sec:LayeredMediaConfig} and the Appendix, the transmission coefficient $T^{(n)}$ is a linear function of $Z_n$; thus, \eqref{eqTZW} may be solved analytically, as done in \cite{PopovPRAppl2019} for the simpler single-layer reflective case considered therein. Here, due to the more elaborate nature of multiple reflections in the transmissive stratified media configuration, we used the implicit equation \eqref{eqTZW} as is, maintaining the formalism's generality.}. This provides the full functional relationship shown in \eqref{eqZkofWk} which is valid for \emph{both} single layer and multilayer meta-atoms.   

Having established \eqref{eqZkofWk} via the procedure associated with \eqref{eqTZW}, we may use the formalism of Section \ref{sec:LayeredMediaConfig} to analyze JC-based configurations in a seamless manner. In particular, \eqref{eqTR} may now be written as
 \begin{equation}
T=T(W_1,W_2,...,W_N) \equiv T(\textbf{W}),
\label{eqReq0Wk}
\end{equation}
where the vector \textbf{W} is composed of the leg lengths $W_n$. The expression in \eqref{eqReq0Wk} establishes the ability of LAYERS to determine the transmission coefficient $T$ of our multi-layered, dual polarized meta-atom, with layer interfaces characterized by JCs with any combination of leg-lengths $W_n$. This capability will now be used to construct a corresponding LUT, allowing synthesis of PCB HMSs based on these constructs.

\subsection{Lookup Table (LUT)}
\label{subsec:LUT}

It will be recalled that to devise a functional HMS, physical realizations of meta-atoms implementing the local scattering responses specified in the \emph{macroscopic} design phase should be employed (see Section \ref{sec:introduction}). To this end, in the current \emph{microscopic} design stage, our aim is to form the relevant LUT, sorting out such suitable meta-atom geometries and relating them to the corresponding scattering parameters. Denoting the complex transmission coefficient $T$ as
\begin{equation}
T=|T|\exp (i\phi), 
\label{eqWT}
\end{equation}
where $\phi$ is the phase of $T$ and can take on values $-180^{\circ}\le \phi \le 180^{\circ}$, the meta-atoms in the LUT will be characterized by their high transmissivity,
 \begin{equation}
|T|=|T(\textbf{W})| \approx 1. 
\label{eqTlarge}
\end{equation}
For populating our LUT, then, we seek vectors $\textbf{W}$ which satisfy \eqref{eqTlarge}. More specifically, we iterate over the entire range of $\phi$, and using our effective semianalytical model (Sections \ref{sec:LayeredMediaConfig} to \ref{sec:DetZkWk}) we seek combinations $\textbf{W}$ that yield the highest $|T|$ for each particular transmission phase. This is one of the core achievements of this work, providing a high-fidelity computationally-viable path to implement the HMS \textit{microscopic} design stage, accounting for near-field interlayer coupling while avoiding full-wave optimization. The resultant LUT would then be composed of Huygens' meta-atom entries, with each meta-atom $\ell$ characterized by the three quantities: $\textbf{W}_\ell$, $|T_\ell|$ and $\phi_\ell$.
As will be demonstrated in Section \ref{sec:ResultsAndDiscussion}, these meta-atoms can be used in the \textit{macroscopic} stage of the design as side-by-side building blocks of a new construct designed to scatter one type of incident wave into a different type of transmitted wave.

\subsection{Frequency Response}
\label{sec:FreqRespTheory}
Once an LUT is obtained as described in Section \ref{subsec:LUT}, we may proceed to devising HMSs for the designated frequency. However, in order to achieve wideband operation -- as required from MSs integrated in communication systems, for instance -- it is critical to identify meta-atom geometries with favorable frequency response, and to select them for our LUT. To accommodate such a feature, we devise in the following a method to easily and efficiently predict the scattering properties of the meta-atoms across a band of frequencies. Indeed, as highlighted in Section \ref{sec:introduction}, this augmented ability is one of the major advances of our extended analytical model over previous work. 

Of course, in the same manner that the extended analytical model, implemented in LAYERS, is capable of evaluating \eqref{eqReq0Wk} for one frequency, it can evaluate it at any different frequency.  But for that different frequency, the original  dependence $Z_n(W_n)$ [Eq. \eqref{eqZkofWk}] (calculated using the original operating frequency) would no longer be valid. That is, in \eqref{eqZkofWk} there is an implicit dependence on the wavelength $\lambda$ that had been suppressed,
 \begin{equation}
Z_n=Z_n(W_n;\lambda),1 \le n \le N,
\label{eqZkofWkf}
\end{equation}
manifesting the inherent dependency of electrodynamic phenomena on the electrical length of the considered configuration. Consequently, if $\lambda$ is changed, a new dependence must be found between $Z_n$ and $W_n$ by re-solving \eqref{eqTZW}, including the full wave CST calculations in that equation, which may incur substantial computational burden (especially for large bandwidths). However, by construction, LAYERS is able to avoid this since it utilizes a unit wavelength, with all structural lengths provided relative to that unit wavelength $\lambda=1$. As a result, instead of redefining the wavelength (or equivalently the frequency), LAYERS maintains the unit wavelength but changes all structural lengths in inverse proportion to the new desired wavelength.  Thus, for example, if $\lambda_a$ were the original actual wavelength for which $T=T_a$ were found, and we wish to use LAYERS to find $T=T_b$ for a different wavelength $\lambda_b$, the new structural lengths would be defined as the old lengths (relative to the original wavelength) multiplied by $\lambda_a/\lambda_b$. This will hold, say, for the parameters $d,w,y_n$, and most interestingly $W_n$.  Since the dielectric constant of the laminates is assumed to be relatively constant across the band, and the wavelength remains unity so that \eqref{eqZkofWkf} still holds, the new values of $W_n=W_{nb}=(\lambda_a/\lambda_b)W_{na}$ may simply be used in the existing relationships $Z_n(W_n)$ to find the new values $Z_n=Z_{nb}$ and corresponding new value of $T=T_b$ for actual wavelength\footnote{This approach implicitly assumes that changes in (i) the copper thickness $t$ and (ii) trace separation $w_s=w$ (Fig. \ref{figCapacitors}b), which are not directly accounted for in LAYERS, but only through the evaluated $Z_n$ (see Appendix), do not greatly affect the distributed impedance (which is considered in the scaling process to be mainly determined by the leg-length $W_n$). The latter assumption (ii) is supported by the fact that the dominant capacitive coupling is only dependent on the ratio between $w$ and $w_s$ \cite{garg2024microstrip}, which indeed remains constant among frequencies; while the former assumption (i) may limit the approximate model's validity to moderate bandwidths. In practice, these assumptions were found to work well for the $10\%$ fractional bandwidth case study presented in Section \ref{sec:ResultsAndDiscussion}.} $\lambda_b$. This method of utilizing $W_n$ as an independent variable in \eqref{eqZkofWk} greatly facilitates the transformation from one frequency to another, equipping LAYERS with a rapid option to evaluate (and subsequently optimize) meta-atom performance across a desired band.

\section{Results and Discussion}
\label{sec:ResultsAndDiscussion}

To demonstrate the validity and efficacy of the methodology proposed in Section \ref{sec:theory} and the corresponding semianalytical model codified in the LAYERS application, we apply it to devise a suitable dual polarized Huygens' meta-atom LUT for the multilayer PCB configuration of Fig. \ref{figIntro}a with $N=5$ interfaces (i.e., four dielectric layers) at $f=20$ GHz. Each laminate is 30 mil thick and its dielectric is characterized by $\varepsilon_n =3$, $\tan \delta_{n}=0.001$ (e.g., similar to Rogers RO3003 or Isola Astra MT77); the period is $d=\lambda/4.9=108.5$ mil; and the JC trace width and separation are $w=w_s=4$ mil (Fig. \ref{figCapacitors}). To bond the PCBs together, we employ a 2-mil compatible prepreg ($\varepsilon =3$, $\tan \delta=0.001$) above interfaces $n=2,3,4$ in Fig. \ref{figIntro}a, so that $y_1-y_2=y_2-y_3=y_3-y_4=32$ mil, while $y_4-y_5=30$ mil.

\subsection{Effective Load Impedance}
\label{sec:DetZkWkResults}
Following Section \ref{sec:DetZkWk} to lay the groundwork for the operation of LAYERS, we begin by considering the relations ${Z_n=Z_n(W_n)}$ between the actual JC geometry characterized by $W_n$ and the effective load impedance $Z_n$ for each of the $N$ interfaces. To this end, we simulate in CST the configuration in Fig. \ref{figfiveinterfaces}b for each of the five interfaces; for each interface $n$, the leg-legnth $W_n$ is swept, considering 30 different values between $W=0$ and $W=80$ mil. Altogether, the full-wave simulations take ${\sim7.5}$ hours to run on a standard PC for the entire configuration (all five interfaces); subsequently, the relations ${Z_n=Z_n(W_n)}$ are readily obtained by solving $\eqref{eqTZW}$ for $Z_n$ (Section \ref{sec:DetZkWk}). 

\begin{figure}[!t]
\centerline{\includegraphics[width=0.8\columnwidth]{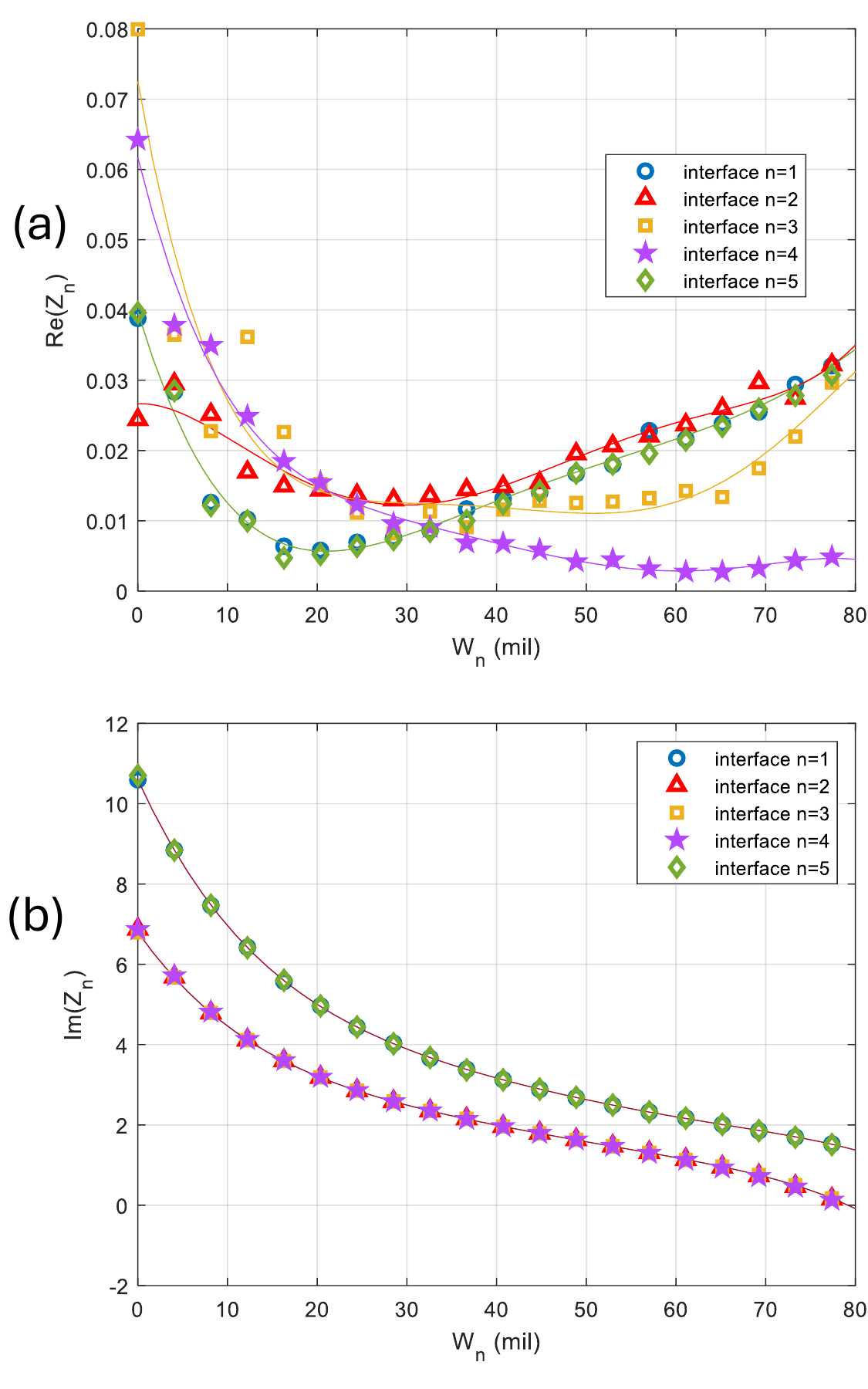}}
\caption{Distributed impedance $Z_n(W_n)$ as a function of the JC leg length $W_n$ for wires at the five interfaces of the MS, utilizing the parameters in Section \ref{sec:DetZkWk}. (a) $\Re(Z_n)$. (b) $\Im(Z_n)$. The data obtained using \eqref{eqTZW} are indicated by the markers. The narrow solid curves are fifth degree polynomial fits for this data that are utilized by LAYERS. $W_n$ values are provided to 80 mil since beyond that value, proper solutions of \eqref{eqTZW} were not found.}
\label{figZWk}
\end{figure}

The results for each interface $n$ are given in Fig. \ref{figZWk}. The real part of the distributed impedance (Fig. \ref{figZWk}a) represents the ohmic losses in the wire.  Since $e^{-i\omega t}$ is the assumed time dependence, the imaginary part (Fig. \ref{figZWk}b) represents capacitive loads since $\Im(Z_n)$ in the figure is positive.

Beyond facilitating the semianalytical LUT construction and frequency response analysis discussed in Sections \ref{subsec:LUT} and \ref{sec:FreqRespTheory}, the extracted effective impedance trends may also provide insights regarding the bahviour of the JC geometries in terms of effective quasistatic loads. Considering Fig. \ref{figZWk}b first, it is apparent that the results for $n=1$ coincide with those for $n=N=5$, and the $n=2,3,4$ results are essentially identical to each other as well. This stems from the capacitive nature of the effective JC reactance, mostly affected by the immediate dielectric environment between the capacitor legs \cite{PfeifferPhysRevAppl2014}. Correspondingly, both external interfaces ($n=1,5$) experience an effective dielectric constant of $\varepsilon_\mathrm{eff}=(1+\varepsilon)/2=2$ \cite{RabinovichIEEETAP2018}; while the internal interfaces ($n=2,3,4$) are well embedded in the substrate, all seeing the same dielectric constant of $\varepsilon=3$. These observations are highlighted by reviewing the effective reactance values in the deeply capacitive regime ($W\rightarrow 0$), where the ratio between the values extracted for the internal and external layers approach $\varepsilon/\varepsilon_\mathrm{eff}=3/2$ as expected. We further note that as the leg-length increases, the capacitance increases (capacitive reactance decreases) while inductive components associated with this conductor elongation become more dominant, driving the effective reactance of Fig. \ref{figZWk}b towards a series RLC resonance (occurring around $W_n\approx 80$ mil for the internal layers $n=2,3,4$), which also contributes to deviation from the aforementioned dielectric constant ratio \cite{diker2026metagratings}.

As per the effective ohmic losses presented in Fig. \ref{figZWk}a, one may observe similar (and rather small) values for all considered interfaces, with identical $\Re{(Z_n)}$ extracted for $n=1$ and $n=5$, stemming from symmetry considerations (minor differences between the effective resistances of $n=2,3,4$ may be attributed to slight deviation from the PCB stack symmetry (Fig. \ref{figfiveinterfaces}a) due to the introduction of bonding prepreg layers). As recently proposed in \cite{diker2026metagratings}, we associate the $\Re{(Z_n(W_n))}$ trends as a function of leg-length mainly with the presence of dielectric loss: the effective resistance increases as $W_n\rightarrow0$ in this deeply capacitive regime as expected for lossy capacitors $\Re{(Z_n)}\propto(\tan\delta)_\mathrm{eff}\Im{(Z_n)}$ when the capacitance goes to zero (reactance goes to infinity) \cite{costa2012closed}; and rises again when $W_n$ approaches the series resonance leg-length (and beyond) due to secondary capacitive coupling between distant JC legs (e.g., the top and bottom legs within the dashed green square of Fig. \ref{figCapacitors}b) becoming more dominant for large values of $W_n$, forming a parallel RLC resonant condition \cite{diker2026metagratings}.

Returning to our main goal here -- establishing the relation \eqref{eqZkofWk} -- each of the data curves in Figs. \ref{figZWk}a and \ref{figZWk}b has been least-square fitted to a fifth degree polynomial, to enhance integration with LAYERS:
 \begin{equation}
\Re(Z_n)=\sum\limits_{\nu=0 }^{5}a_\nu W_n^\nu,\,\,\,\,\Im(Z_n)=\sum\limits_{\nu=0 }^{5}b_\nu W_n^\nu,
\label{eqPolynom}
\end{equation}
where the coefficients $a_\nu,b_\nu$ are found through the standard MATLAB function \texttt{polyfit} and input to the code. In LAYERS, \eqref{eqPolynom} is used in \eqref{eqTR} to produce the transmission coefficient as given in \eqref{eqReq0Wk} for any valid leg-length combination $\textbf{W}$, enabling -- as shall be promptly exemplified -- effective exploration of the design space for identifying optimal meta-atoms for populating our LUT.

\subsection{Microscopic Design: Building the Lookup Table}
\label{sec:buildingLookupTable}
Although we have shown that we may obtain the transmission coefficient $T$ across the array of meta-atoms for a given set $\textbf{W}$ of JC leg-lengths $W_n$, we are interested in solving the inverse problem: as stated in Section \ref{subsec:LUT}, we seek sets  $\textbf{W}_\ell$ for which $|T_\ell(\textbf{W}_\ell)|\approx 1$, with each set $\textbf{W}_\ell$ producing a different phase $\phi_\ell$ of $T_\ell$. Even though it is possible to find such sets by applying optimization routines to the  LAYERS model, this often requires multiple user interactions to assure success and proper coverage of $\phi$.  Instead, we employ a less sophisticated but conceptually simpler procedure for finding the $\textbf{W}_\ell$ of interest: we perform "exhaustive" sampling of the length $W_n$ at each of the $N$ interfaces.  This sampling is implemented at equal intervals of $\Delta_W$ mil, and the $T_\ell$ for each of these combinations is determined by LAYERS. These values of $T_\ell$ will cover the entire unit circle $|T|\le 1$ in the complex $T$ plane, but, as shown below, the desired points $|T_\ell|\approx 1$  can be easily retrieved. This exhaustive sampling can be facilitated with little or no user interaction and can be easily manipulated programmatically. Focusing on HMSs, which should feature a symmteric meta-atom configuration \cite{MonticonePRL2013,epstein2016huygens,EpsteinIEEETAP2016}, the implementation can be made more efficient by realizing that, except for the presence of the very thin bond between the laminates, the dielectric geometry assumed in Section \ref{sec:DetZkWk} is symmetrical about the center ($n=3$) interface.
Thus, to reflect the required HMS symmetry, we shall demand that the $W_n$ be symmetric, so that in our exhaustive sampling, we can set $W_1=W_N$, $W_2=W_{N-1}$,....  

\begin{figure}[!t]
\centerline{\includegraphics[width=0.8\columnwidth]{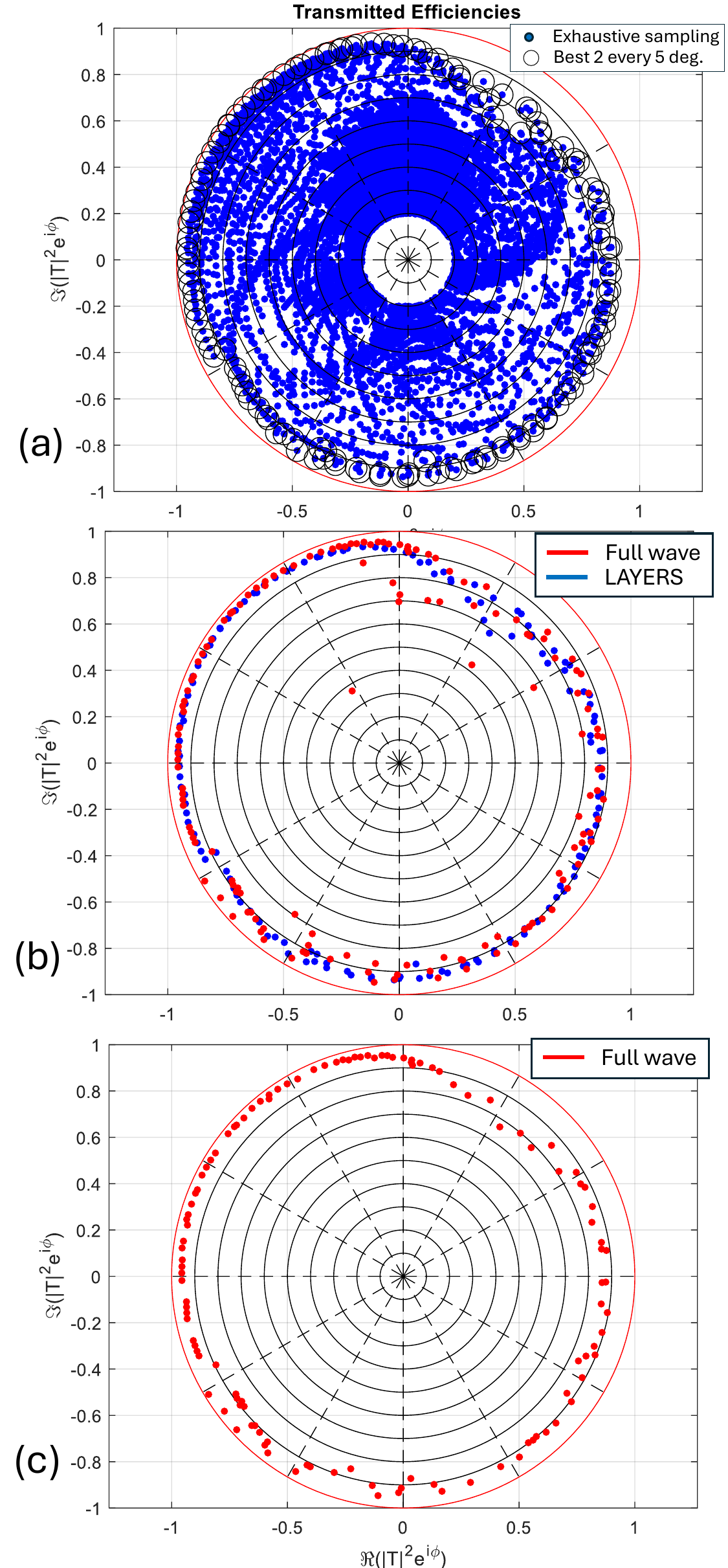}}
\caption{Values of $|T_\ell|^2e^{i\phi_\ell}$ located in a complex unit circle for the meta-atom defined in Section \ref{sec:DetZkWk}. The dashed radials are drawn every $30^{\circ}$, and the solid circles are drawn at intervals of 0.1. (a) Results as calculated by LAYERS of an exhaustive sampling, $\Delta_W=2$ mil, with only $|T_\ell|^2>0.2$ displayed.  The circled dots represent the two largest values of $|T_\ell|^2$ (i.e. the two closest values to the red circle $|T|^2=1$) in every phase interval of $5^{\circ}$.  (b) The two largest LAYERS-computed values of $|T_\ell|^2$ in every phase interval of $5^{\circ}$ that were circled in (a) (blue dots), and the corresponding full-wave-computed values of $|T_\ell|^2e^{i\phi_\ell}$ for the same sets $\textbf{W}_\ell$.  (c) The final lookup table consisting of the full-wave-calculated values of $|T_\ell|^2e^{i\phi_\ell}$ of (b), but with smaller values of  $|T_\ell|^2$ removed.  }
\label{figZExh1}
\end{figure}

Consequently, for the case at hand where $N=5$, only $W_1$, $W_2$ and $W_3$ need be varied. Using a sampling interval $\Delta_W=2$ mil, and recalling from Fig. \ref{figZWk} that $W_n$ extends from 0 to 80 mil, there are 41 intervals at each interface, creating a total of $41^3$ samples of $\textbf{W}_\ell$ for the three independent interfaces. LAYERS required less than 30 minutes (using a standard PC) to determine the $T_\ell$ corresponding to these samples. Fig. \ref{figZExh1}a shows these values for which $|T|^2>0.2$, where $|T|^2$ is the power coupling efficiency from the incident wave to the transmitted wave.  Each blue dot in Fig. \ref{figZExh1}a represents a complex number $|T_\ell|^2e^{i\phi_\ell}$ which corresponds to a set $\textbf{W}_\ell$ of five JC leg lengths.

	Since we are interested in configurations with maximum efficiency (i.e. $|T_\ell|^2$ closest to 1), and for different phases $\phi$, we would choose the results in Fig. \ref{figZExh1}a that lie along the outer periphery of the cloud of blue dots. The two largest values of $|T_\ell|^2$ within each  $5^{\circ}$ interval are those circled in the figure. This number of solutions was chosen because it is about twice the number required for a reasonable LUT, thereby allowing the possibility of eliminating several of them for reasons that will become clear below. 
	
Despite the rigorous incorporation of near- and far-field interlayer and intralayer coupling within the meta-atom PCB stack (see Section \ref{sec:theory} and Appendix), the uniformly distributed effective impedance model used in our analysis scheme restricts its validity to the dipole moment contribution; higher-order multipoles possibly stemming from the exact current profiles on the JCs are neglected. Correspondingly, we may expect some deviations of	
the predicted values of  $|T_\ell|^2$ and $\phi_\ell$ in Fig. \ref{figZExh1}a with respect to the ground truth. Therefore, the final step in the microscopic design stage of the methodology presented herein includes full-wave verification of the candidate meta-atoms before officially registering their parameters in the LUT. 
	
	Specifically, we employ a full-wave solver to determine the $|T_\ell|^2$, $\phi_\ell$ for the same set $\textbf{W}_\ell$ which had produced the circled blue dots. These full-wave (CST) results are shown in Fig. \ref{figZExh1}b as red dots, along with LAYERS (Matlab) calculations shown as the same blue dots (without their circles) using the same $\textbf{W}_\ell$. For the most part, the full-wave results (red) lie along the same circumference as the analytical results (blue), but with some displaying smaller magnitudes.  Since only large values of $|T_\ell|^2$ should be included in the LUT, and since a larger-than-necessary number of solutions was processed, it is possible to eliminate the small-magnitude solutions when this does not adversely affect the $\phi$-density of the solutions. Following this removal of the small $|T_\ell|^2$, the LUT of Fig. \ref{figZExh1}c will result.  It will be noticed that only full-wave results are included in this LUT, there no longer being a need for the analytical results; the ultimately produced LUT is, thus, of absolute (CST-generated) accuracy. It should be emphasized, however, that these full-wave results could not have been achieved this efficiently without LAYERS providing the sets $\textbf{W}_\ell$. In this sense, the method for defining the LUT is referred to as "semianalytical" in spite of LAYERS itself being analytical. In fact, these full-wave calculations of $|T_\ell|^2$ and $\phi_\ell$ for all entries $\textbf{W}_\ell$ in the LUT is the most time-consuming portion of the presented microscopic design procedure, providing strong motivation for further devising means to alleviate this need. Indeed, employing machine learning methods to this end is the focus of Part II of this two-part compilation \cite{Nissan2026}, laying out the development of such a scheme and demonstrating its efficacy.

\subsection{Macroscopic Design: Utilizing the Lookup Table}
\label{subsec:macroscopic_design}
\begin{figure}[!t]
\centerline{\includegraphics[width=0.8\columnwidth]{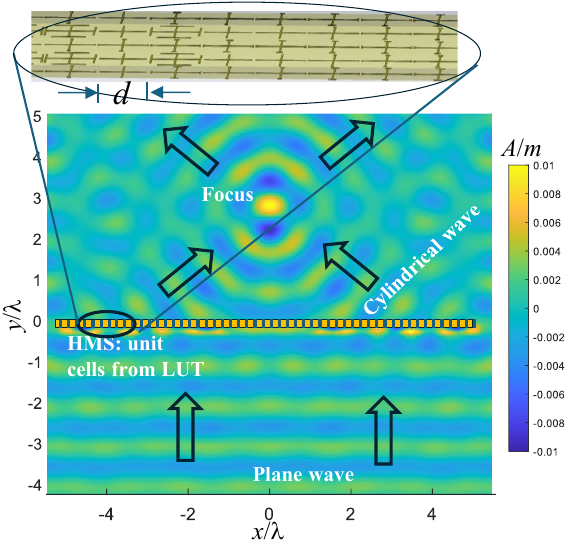}}
\caption{Full-wave results of $\Re(H_{z})$ for a TM plane wave incident on the HMS from below, that is transformed by the HMS to a cylindrical wave in the region $y>0$. The HMS is about $10\lambda$ in length, is periodic in the $z$-direction, and its upper face coincides with the $x$-$z$ plane. Its center is at $x$=0, and it is composed of 51 unit cells from the LUT (Fig. \ref{figZExh1}c) chosen to satisfy \eqref{eqphin} and \eqref{eqTn}. The E-field of the incident plane wave is 1 volt per meter.  The HMS focuses the wave to a point which lies at a distance $y_c=3\lambda$ from the HMS. An enlarged view of 8 unit cells of the HMS is shown above the color image. }
\label{figHfield}
\end{figure}

Now that the LUT has been created (Fig. \ref{figZExh1}c), its entries will be used to devise a HMS to transform one type of incident field to a different type of transmitted field. In particular, as a representative case study for this macroscopic design, we aim at synthesizing a metalens, converting a plane wave propagating normal to the HMS into a cylindrical wave, the $z$-directed axis of which is parallel to the HMS and is located at $x=0$, $y=y_c>0$, so that the center of the wave (focus of the metalens) is a distance $y_c$ from the surface\footnote{The same methodology presented here can be used to design a standard 3D dual-polarized metalens, converting a plane wave into a spherical wave. We restrict ourselves here to a 2D scenario $\partial/\partial z=0$ only due to simulation resource constraints, preventing us from showcasing the full 3D performance of the finalized metalens.} (Fig. \ref{figHfield}).

To facilitate the macroscopic design \cite{epstein2016huygens}, we need to identify the required local meta-atom properties as a function of their location along the HMS. For a wave converging on the cylinder axis, its phase to an additive constant along the top of the HMS will be $\phi^+(x)=2\pi \sqrt{x^2+y_c^2}/\lambda$. Along the bottom of the HMS, the phase will be that of a plane wave propagating normal to the surface, so that $\phi^-(x)$ would be constant.  It is assumed that this constant is the value of the phase at $x=0$ on the upper surface, so that $\phi^-(x)=2\pi y_c/\lambda$.
The HMS is thus constructed in the $x$-$z$ plane by placing unit cells from the lookup table at intervals $d$ along the $x$-axis \cite{epstein2016huygens, Liu20Apertures, Chen2023Microwave}. Each cell $\nu$ with center at $x_\nu$ will be chosen so that its phase $\phi_\nu$ is closest to the phase difference between the top and the bottom of the HMS: 
 \begin{equation}
\phi_\nu=\phi^-(x_\nu)-\phi^+(x_\nu).
\label{eqphin}
\end{equation}
As for the macroscopic design requirements for $|T(x)|$, we follow here the conventional metalens (phase-gradient) synthesis approach and aim at maximizing the transmission coefficient magnitude at all points along the HMS \cite{Liu20Apertures, Chen2023Microwave}. While this may incur some reflection loss due to wave impedance mismatch between incident and departing "rays" \cite{epstein2016huygens,WongAWPL2016,EpsteinIEEETAP2016,kang2020efficient}, for metalens applications with moderate $f/D$ (focal distance to diameter) ratios, the overall efficiency should not be greatly affected. Thus, we employ here this approach as well, favoring design simplicity over absolute performance, striving to position at each $x_\nu$ a meta-atom with the highest possible local transmission efficiency, namely
 \begin{equation}
|T_\nu|=1.
\label{eqTn}
\end{equation}
The requirements \eqref{eqphin} and \eqref{eqTn} are thus completely compatible with the criteria on which the LUT in Fig. \ref{figZExh1}c are based. That LUT, devised by our LAYERS application, therefore represents the microscopic design resource for constructing the current macroscopic HMS layout. More specifically, for each cell $\nu$, we identify in the LUT (Fig. \ref{figZExh1}c) the most suitable meta-atom configuration -- corresponding to the fabrication-ready realistic PCB metallodielectric stack with the selected $\textbf{W}_\nu$ leg-length combination -- and position it at $x_\nu$ to form a segment of the metalens.

\begin{figure}[!t]
\centerline{\includegraphics[width=0.8\columnwidth]{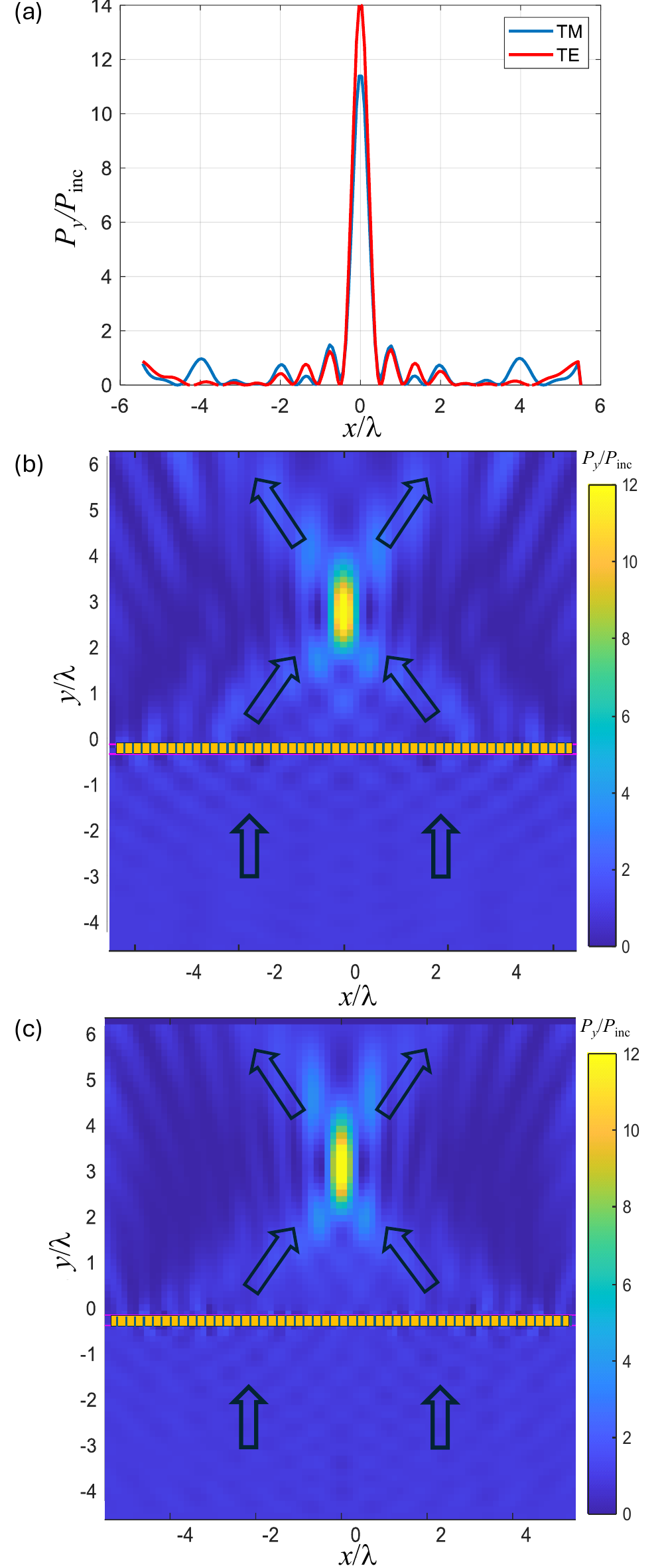}}
\caption{Full wave (CST) results of a plane wave incident from below on the same HMS as in Fig. \ref{figHfield}, and focusing its power to a point $y=y_c=3\lambda$ above the HMS. (a) The $y$-component $P_y$ of the power flow relative to the incident power flow $P_{\text{inc}}$ at $y=y_c=3$ as a function of $x/\lambda$ for TE and TM plane wave incidence. (b) The power flow in the $x$-$y$ plane for the TM case.  (c) The power flow in the $x$-$y$ plane for the TE case.} 
\label{figHfield2}
\end{figure}

Applying the outlined macroscopic design procedure to a $10\lambda$-long metalens with $y_c=3\lambda$ using the LUT of Section \ref{sec:buildingLookupTable} (Fig. \ref{figZExh1}c), we proceed to simulate its illumination by a TM polarized plane wave normally incident on it from below. Fig. \ref{figHfield} displays the full-wave-calculated H-field, displaying the transformation by the metalens of the incident plane wave into a "cylindrical" wave above the HMS propagating towards the focal point at $x=0$, $y=y_c$.  The relative power flow along the line through this focus and parallel to the $x$-axis is shown in Fig. \ref{figHfield2}a for both TE and TM incident plane waves, highlighting the successful dual-polarized operation of our design. To complement this, Figs. \ref{figHfield2}b and \ref{figHfield2}c display the relative power flow within the $x$-$y$ plane for TM and TE incidence, respectively. Only slight differences, consistent with Fig. \ref{figHfield2}a, are apparent between these figures. An analysis of the full-wave solutions indicates that, for TM, about 80.0\% of the incident power of the plane wave reaches the focal plane at $y=y_c=3\lambda$, as determined by integrating the normal component of the power flow vector over the metalens aperture range in this plane. About 5.8\% of the incident power is scattered back into the incident region, and the remaining 14.2\% is attributed to losses. For TE-polarized illumination, about 83.2\% of the incident power of the plane wave is transferred to the focal plane, 4.8\% is scattered back into the incident region, and 12.0\% is attributed to losses. The corresponding full width at half maximum (FWHM) values deduced from Fig. \ref{figHfield2}a for the TM and TE polarizations, respectively, are $0.462\lambda$ and $0.460\lambda$, indicating the proper (dual-polarized) formation of the focus as prescribed.

Figures \ref{figHfield} and \ref{figHfield2} and their attendant analyses demonstrate the success of the LUT entries in creating realistic HMS constructs that provide efficient control of waves transmitted through them, and consequently the success of our proposed semianalytical methodology. Overall, relying on a rigorous analytical model considering practical PCB-compatible configurations and including relevant conductor and dielectric loss, the developed scheme and its codified version LAYERS form an effective engineering tool for on-demand synthesis of versatile fabrication-ready dual-polarized HMSs.

\subsection{Frequency Response}
\label{sec:FreqRespResults}
The results discussed thus far in Section \ref{sec:ResultsAndDiscussion} have been valid for a single frequency (20 GHz in our case). Nonetheless, as noted in Section \ref{sec:FreqRespTheory}, many applications require wideband operation, which, in turn, requires the creation of a LUT with favorable transmission $T$ properties over a spectrum of frequencies. It will be recalled that in Section \ref{sec:FreqRespTheory} we have developed a judicious analytical scaling rule, by which LAYERS can be used to predict the frequency response of a given JC-based meta-atom of the type considered herein (Fig. \ref{figIntro}a). Thus, as a first step of validation, we employ the methods presented therein to the single-frequency LUT entries of Fig. \ref{figZExh1}c, and compare the predicted frequency-dependent meta-atom scattering properties to the CST simulated values. 

\begin{figure}[!t]
\centerline{\includegraphics[width=0.8\columnwidth]{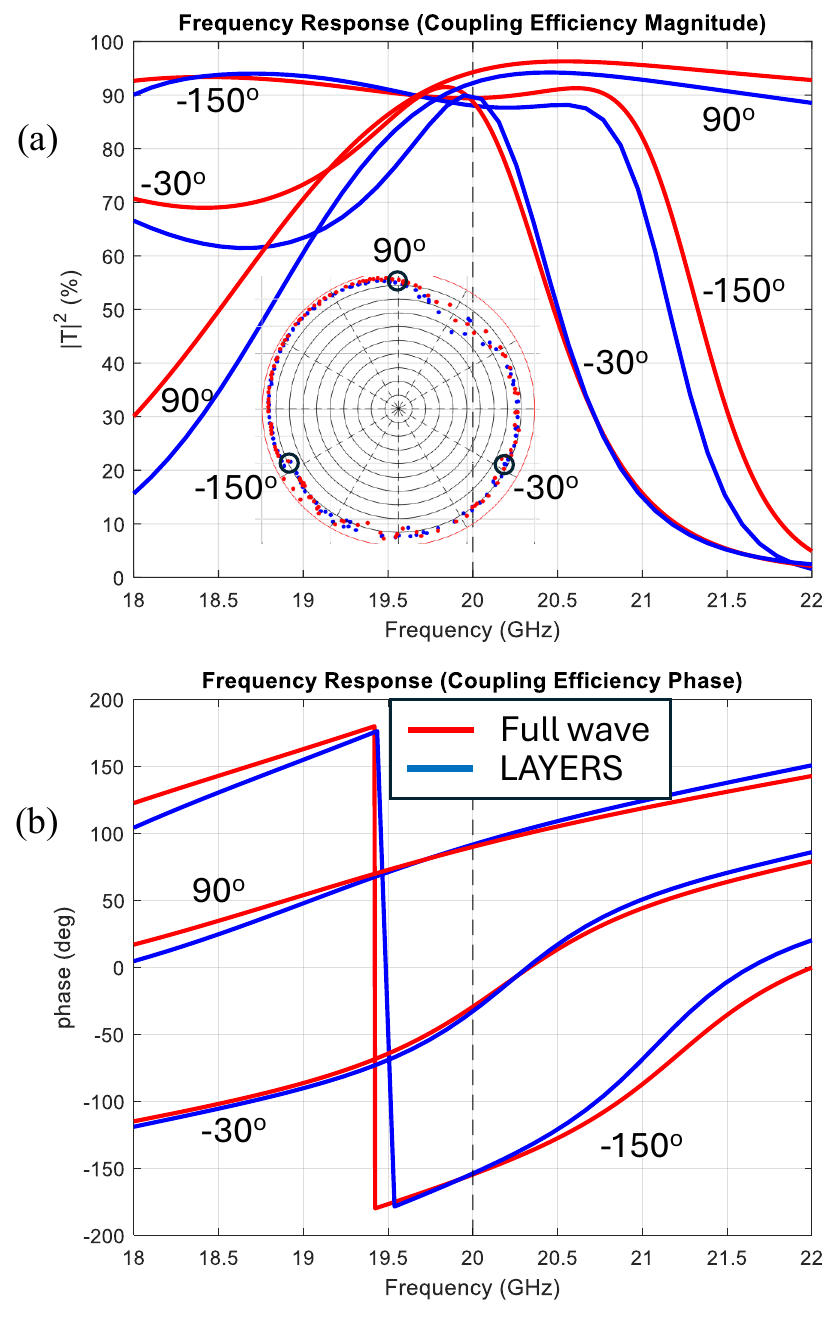}}
\caption{The frequency response for three sets of $\textbf{W}$ from the lookup table of Fig. \ref{figZExh1}, as computed analytically by LAYERS (blue curves) and numerically by a full-wave solver (red curves). The three sets are near coupling efficiency phases $\phi=-150^{\circ}$, $\phi=-30^{\circ}$ and $\phi=90^{\circ}$, as indicated in the inset of (a). These correspond to $W=\{16,70,10,70,16\}$ mil, $W=\{14,8,10,8,14\}$ mil and $W=\{0,58,2,58,0\}$ mil, respectively. The 20 GHz nominal frequency is  shown by a dashed vertical line, and the frequency range is 18 to 22 GHz. (a) Coupling efficiency magnitude vs. frequency. (b) Coupling efficiency phase vs. frequency.}
\label{figFreqResp1}
\end{figure}

Correspondingly, a comparison of $|T|^2e^{i\phi}$ using LAYERS and using full-wave calculations for a span of frequencies is given in Fig. \ref{figFreqResp1}. There, the coupling efficiency from the incident wave to the transmitted wave is shown for structures from three different phase regions of the LUT of Fig. \ref{figZExh1}b: $\phi\approx-150^{\circ}$, $\phi\approx-30^{\circ}$ and $\phi\approx 90^{\circ}$ (see inset of Fig. \ref{figFreqResp1}a), considering three unit cells which demonstrated good correspondence between LAYERS and CST at the nominal 20 GHz frequency. Separate plots are provided for the magnitude (Fig. \ref{figFreqResp1}a) and for the phase (Fig. \ref{figFreqResp1}b) as functions of frequency over a $10\%$ fractional bandwidth, showcasing good agreement between semianalytical predictions and full-wave results for both.  

\begin{figure}[!t]
\centerline{\includegraphics[width=0.8\columnwidth]{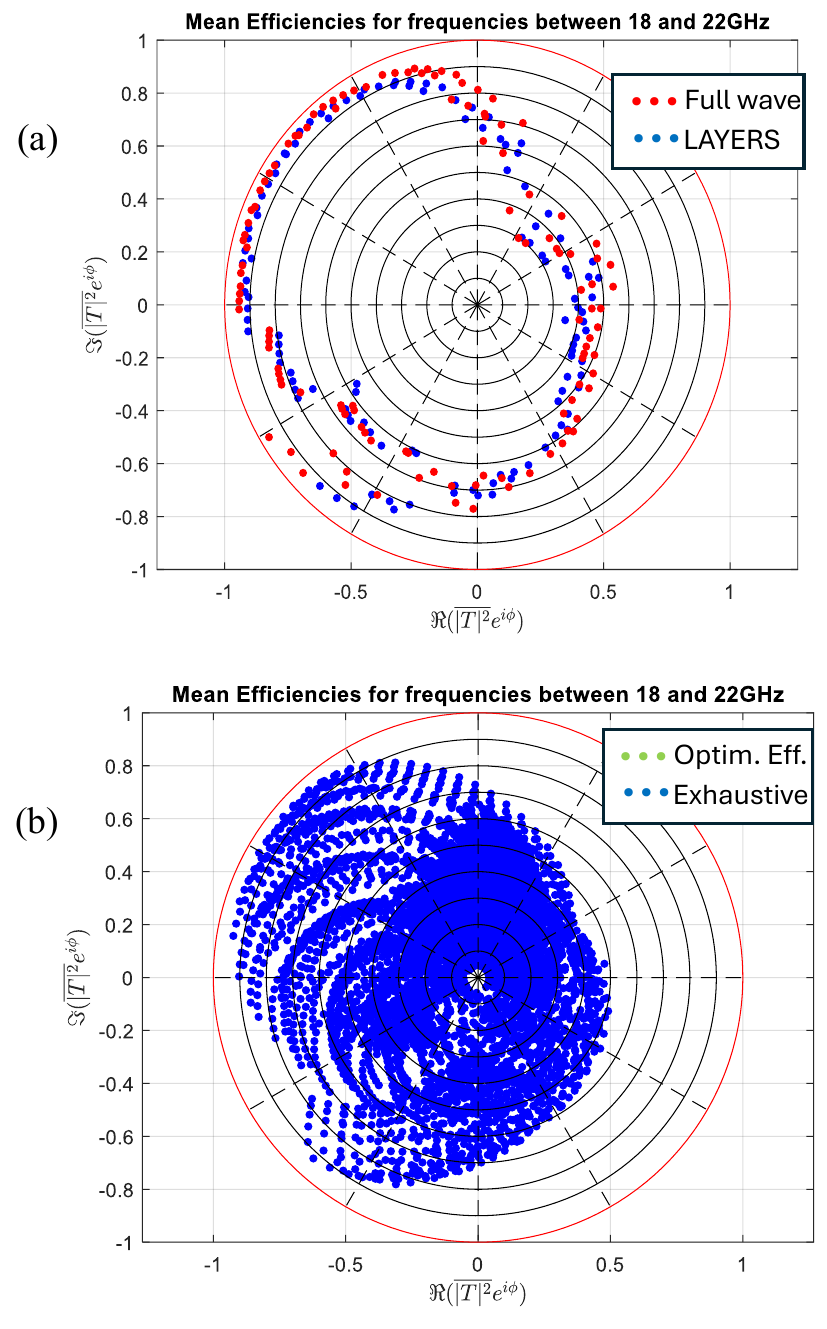}}
\caption{Mean efficiencies as defined in \eqref{eqMeanEff1} for the $\textbf{W}$ configurations to which Fig. \ref{figZExh1} applies, $f_1=18$ GHz, $f_2=22$ GHz. (a) LAYERS and full-wave calculations for the $\textbf{W}$ configurations to which Figs. \ref{figZExh1}b and (c) apply. These were found as the configurations which optimize the efficiency $T$ at the nominal frequency $f_0=20$ GHz. (b) LAYERS calculations for the $\textbf{W}$ configurations to which Fig. \ref{figZExh1}a applies, which represent the entire exhaustive search.} 
\label{figFreqResp2}
\end{figure}

Beyond this basic validation, it would be insightful to demonstrate the utilization of our developed frequency response predictive tool embedded in LAYERS for further performance analysis and enhancement. To facilitate this, we wish to define a figure of merit that we may associate with each of the curves in Fig. \ref{figFreqResp1}, which would provide an effective evaluation of the suitability of the corresponding meta-atom for wideband HMS applications. In the frame of our case study, we choose to this end to use the meta-atom mean transmission efficiency over this frequency range $\overline{|T|^2}$ as such a simple figure of merit, defined as
 \begin{equation}
\overline{|T|^2}=\frac{1}{f_2-f_1}\int_{f_1}^{f_2} |T(\textbf{W};\lambda=c/f)|^2 \,df,
\label{eqMeanEff1}
\end{equation}
where the agrument of $T$ has been generalized from that used in \eqref{eqReq0Wk} to include $\lambda$, as in \eqref{eqZkofWkf}. In \eqref{eqMeanEff1}, $f$ is the frequency in GHz, and $f_1$ and $f_2$ are its initial and final values (18 and 22 GHz in Fig. \ref{figFreqResp1}). With this definition, it is possible to evaluate the mean efficiency of each of the structures in the lookup table of Figs. \ref{figZExh1}b and c. Since each structure can be characterized by its transmitted efficiency phase at the nominal frequency $f_0$ (20 GHz in Fig. \ref{figFreqResp1}), the mean efficiency of all the structures in the lookup table can be represented as points in a "mean efficiency lookup table" in the same manner as in Fig. \ref{figZExh1}, but with the radial component representing the mean efficiency $\overline{|T|^2}$ over the frequency range $[f_1$  $f_2]$ rather than the efficiency $|T|^2$ at $f_0$. This is displayed in Fig. \ref{figFreqResp2}a as calculated both by LAYERS and by full-wave computations.   

It is clear from Fig. \ref{figFreqResp2}a that there is a region $90^{\circ}<\phi<180^{\circ}$ in which $\overline{|T|^2}$ has values similar to $|T|^2$, but for most of the remaining phase angles this is not the case. Presumably, these latter results would adversely affect the field quality displayed in Figs. \ref{figHfield} and \ref{figHfield2} if these are employed at other than the nominal frequency. It should be recalled, however, that the points in Fig. \ref{figFreqResp2}a were based on sets of $\textbf{W}$ in Fig. \ref{figZExh1} that optimize $|T|^2$, not the frequency-averaged mean efficiency $\overline{|T|^2}$. The question, then, is: Can other points (i.e. sets of $\textbf{W}$) be found that would optimize $\overline{|T|^2}$, promoting effective operation across the entire band and not only at the nominal frequency? This may be determined by assessing the values of $\overline{|T|^2}$ for the various meta-atoms configurations considered in the exhaustive search, i. e. by applying \eqref{eqMeanEff1} to the sets $\textbf{W}$ associated with Fig. \ref{figZExh1}a. This is shown in Fig. \ref{figFreqResp2}b, where the optimum values of $\overline{|T|^2}$ would lie on or near the perimeter of the cloud of blue dots. But even these optimum values leave much to be desired, since there are phases for which they are as low as 0.5. Correspondingly, as a final example of the versatility of LAYERS, we shall now show that it is possible to improve these optimum values of $\overline{|T|^2}$ via the devised semianalytical approach.

\begin{figure}[!t]
\centerline{\includegraphics[width=0.8\columnwidth]{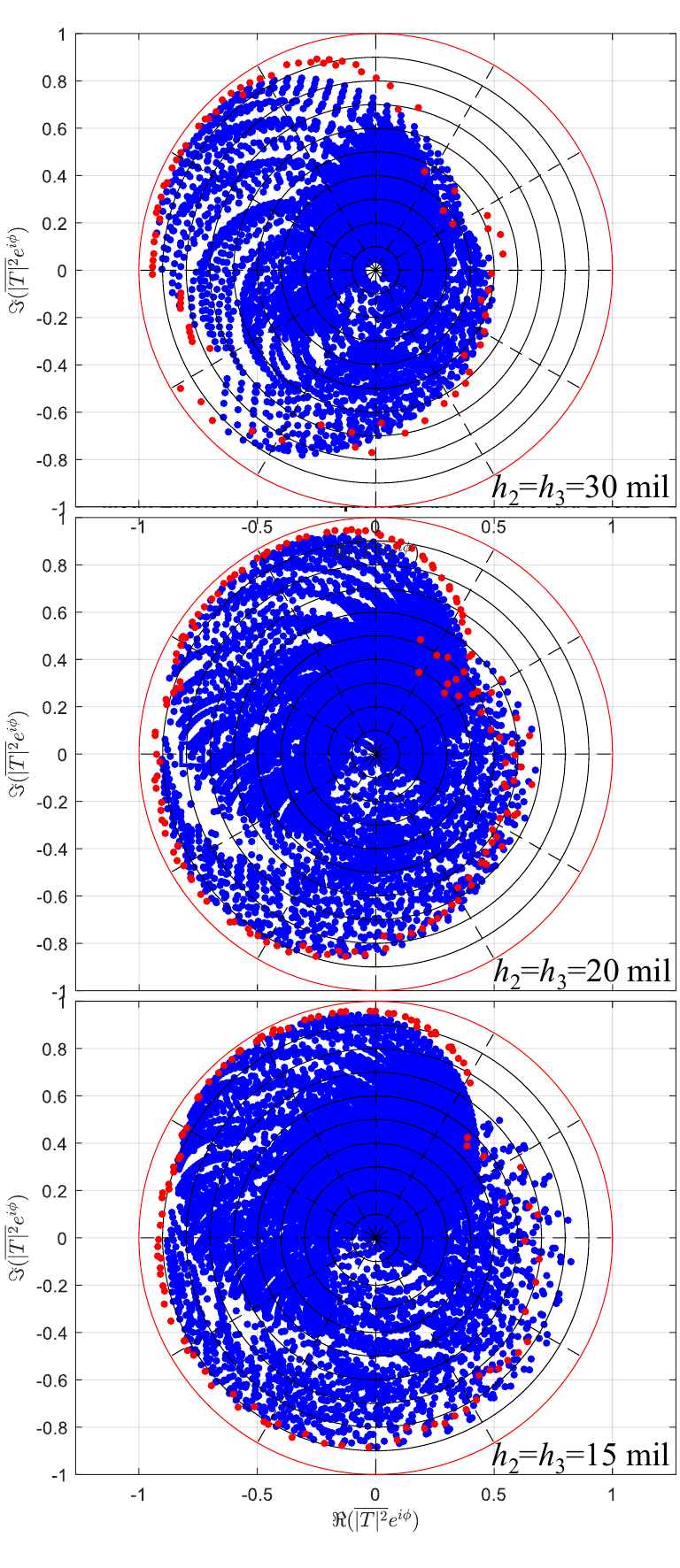}}
\caption{The LAYERS-calculated mean efficiencies (blue dots) over a frequency band [18 22] GHz from exhaustive search configurations obtained for three different meta-atom structures: $h_2=h_3=30$ mil, $h_2=h_3=20$ mil and $h_2=h_3=15$ mil. The red dots indicate the CST-calculated mean efficiencies for meta-atom configurations corresponding to the blue dots along the perimeter of each respective blue-dot-cloud.}   
\label{figMeanEffClouds}
\end{figure}
\begin{figure}[!t]
\centerline{\includegraphics[width=0.7\columnwidth]{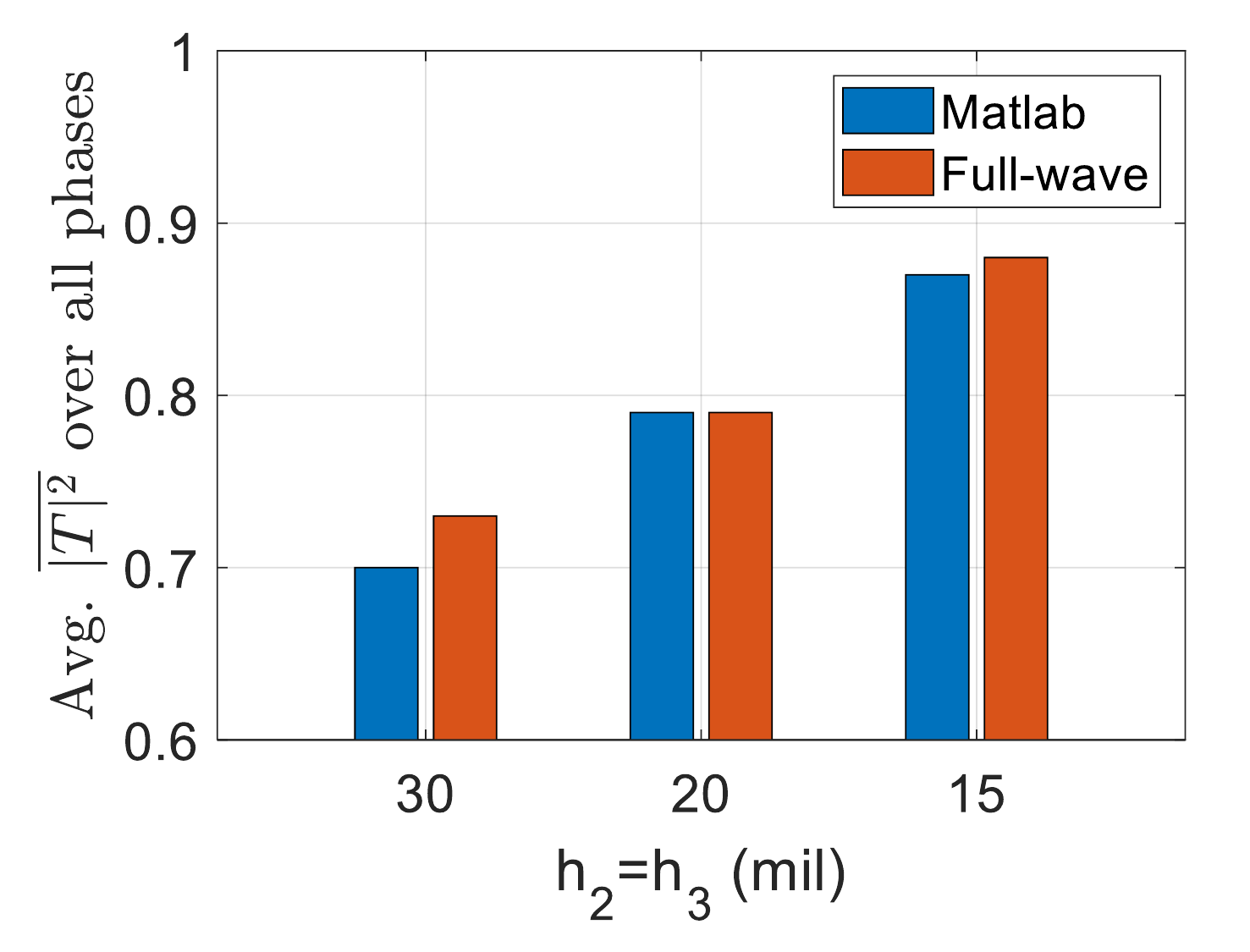}}
\caption{Comparison of mean efficiencies averaged over all phases for three different values of $h_2=h_3$, as calculated by full-wave methods and by LAYERS.}   
\label{figMeanEffBarGraph}
\end{figure}

Our improvement procedure starts by noting that the exhaustive search results shown in Fig. \ref{figFreqResp2}b were obtained for a specific predetermined dielectric stack; in particular, for the case in which the laminate thickness of each layer was 30 mil. However, since the analytical model underlying LAYERS can be rigorously applied to any stratified media configuration (see Appendix), we may readily consider the layer thicknesses as an additional degree of freedom, which may assist enhancing other meta-atom attributes, e.g. the considered frequency response. Consequently, denoting the thickness of the laminate of layer $n$ as $h_n$ (see Fig. \ref{figIntro}), we will attempt to achieve this by considering two other values of the internal layers $h_2=h_3$: 20 mil and 15 mil.  

The exhaustive search "cloud" for the corresponding LAYERS-calculated mean efficiencies is given in Fig. \ref{figMeanEffClouds} for the various values of $h_2=h_3$. The unit circle region that is entirely empty of blue dots when $h_2=30$ mil becomes more populated as $h_2$ decreases, so that the cloud perimeter becomes closer to the boundary of the unit circle, and the efficiencies improve accordingly. A measure of the quality of a mean efficiency LUT is the average value $\mathcal{E}$ of the mean efficiency $\overline{|T|^2}(\phi)$ of the LUT over \emph{all} phases $\phi$:
 \begin{equation}
\mathcal{E}=\frac{1}{2\pi} \int_{-\pi}^{\pi}\overline{|T|^2}(\phi)d\phi,
\label{eqMeanMean}
\end{equation}
where the LUT is defined as the meta-atom configurations along the perimeter of the cloud. Using \eqref{eqMeanMean}, the average mean efficiencies for the three structures considered in Fig. \ref{figMeanEffClouds} are summarized in Fig. \ref{figMeanEffBarGraph}. In that latter figure, it is seen that an improvement of over 15\% can be realized in the average mean efficiency by manipulating a parameter (in this case a laminate thickness) that is simple to change in LAYERS. These results, verified by the full-wave data presented in Figs. \ref{figMeanEffClouds} and \ref{figMeanEffBarGraph}, complete the exposition and validation of our semianalytical Huygens' meta-atom synthesis scheme, demonstrating its versatility and effectiveness in producing realistic PCB-compatible designs for transmissive field manipulation -- both for the central operating frequency as well as over a moderate bandwidth.

\section{Conclusion}

A semianalytical scheme has been presented for designing transmissive dual polarized multilayer meta-atoms without requiring full-wave optimization. This was accomplished by adapting analytical models, originally intended for synthesis of loaded-wire metagratings through rigorous consideration of intralayer and interlayer coupling, to provide reliable predictions of the scattered fields off periodically arranged cascades of JC shaped conductors. Relying on this framework, codified into an HMS microscopic design application called LAYERS, LUTs of realistic PCB-layered unit cells containing these JC etchings may be created, each configuration providing high transmission magnitude at a different phase. Notably, using the implemented wavelength-scaled analysis approach, the LUT entries can be categorized not only by efficiency at a nominal frequency, but also by mean efficiency over a band of frequencies. Utilizing the resultant LUT for case study at K band, we demonstrated the successful macroscopic design of a HMS metalens, forming very good focus properties for both polarizations, as verified in full-wave simulations.  

While this establishes the efficacy and versatility of the developed semianalyitcal methodology and corresponding software tool (LAYERS) in generating efficient fabrication-ready PCB Huygens' meta-atoms on demand, a final full-wave validation of the produced LUT was integrated as part of the procedure, required to address regions of limited accuracy due to the underlying dipole approximations. The machine learning approach devised in the companion paper (Part II) make use of the advances presented herein to alleviate this hurdle: combining rapid training via semianlaytical LAYERS with minimal full-wave fine tuning, the hybrid MetaMamba presented therein enables inverse design of Huygens' meta-atoms which is resource-efficient -- avoiding the enormous full-wave effort typically needed for training -- while at the same time reaches near-ground-truth accuracy throughout the design space.
Together, Parts I and II provide an efficient, open‑source \cite{github} toolkit for rapid, accurate design of fabrication‑ready dual‑polarized transmissive Huygens’ meta‑atoms and metasurfaces.

\appendix
\setcounter{equation}{0}\renewcommand\theequation{A\arabic{equation}}

The mathematical basis of LAYERS for the most part has been presented previously \cite{EpsteinPhysRevAppl2017,RabinovichIEEETAP2018,RabinovichIEEETAP2020}. For completeness, it is reviewed now with minor modifications. It computes the transmission coefficient of a multilayered periodic array of meta-atoms by solving a system of linear equations formed from the electric and magnetic field boundary conditions at the interfaces between the layers, and satisfying Ohm's law on the wires. It accomplishes this while fully accounting for evanescent components of the Floquet spectrum.

\begin{figure}[!t]
\centerline{\includegraphics[width=0.6\columnwidth]{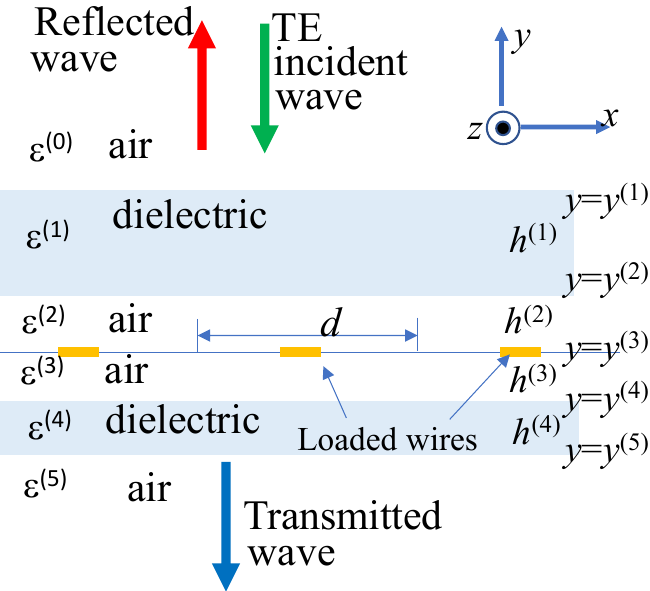}}
\caption{The layered configuration used to derive the basis of the analytical LAYERS model. The incident wave is from above. The reflected and transmitted waves shown assume that the period $d$ is sufficiently small that only a single propagating mode is scattered.} 
\label{figAppend}
\end{figure}

Although LAYERS is applicable to any number of laminate layers (see Fig. \ref{figIntro}), only the subset of layers shown in Fig. \ref{figAppend} will be considered for this description, with the carryover to the general case being straightforward.  Note that Fig. \ref{figAppend} is a bit different from Fig. \ref{figIntro}. In Fig. \ref{figIntro}, the wires were adjacent to the dielectric, whereas in Fig. \ref{figAppend} air space has been inserted between the wires and the dielectrics. This is to enable a unified solution for the electromagnetic behavior of the wires in air; the width of the air space will later be contracted to zero.

Consider, then, the configuration in Fig. \ref{figAppend} consisting of dielectric layers and a periodic array of capacitively loaded wires with period $d$. The dielectric constant of layer $m$ is $\varepsilon^{(m)}$ and its thickness is $h^{(m)}=y^{(m+1)}-y^{(m)}$. The configuration is invariant in the $z$-direction, and an $e^{-i\omega t}$ time dependence is suppressed.  Assuming TE polarization, the E-field will be in the $z$-direction, and in each region $m$, can be written as the sum of Floquet modes:
 \begin{equation}
E_z^{(m)}(x,y)=E_{z,\text{inc}}\delta_{m0}+\sum\limits_{p=-\infty }^{\infty }e^{ik_{xp}x}E_p^{(m)}(y),
\label{eqAppend1}
\end{equation}
where
\begin{equation}
E_{z,\text{inc}}=E_0e^{ik_{x0}x}e^{-ik_{y0}^{(0)}(y-y^{(1)})},
\label{eqAppend2}
\end{equation}
\begin{multline}
E_p^{(m)}(y)=A_p^{(m)}e^{-ik_{yp}^{(m)}(y-y^{(m)})}+B_p^{(m)}e^{ik_{yp}^{(m)}(y-y^{(m)})},\\ 1\le m \le M,
\label{eqAppend3}
\end{multline}
\begin{equation}
E_p^{(m)}(y)=B_p^{(m)}e^{ik_{yp}^{(m)}(y-y^{(1)})},m=0,
\label{eqAppend4}
\end{equation}
\begin{equation}
k_{xp}=k\sin \theta _{\text{inc}}+2p\pi /d,  k_{yp}^{(m)}=\sqrt{{k^{(m)}}^{2}-k_{xp}^2},
\label{eqAppend5}
\end{equation}
$m$ is the sequential index of the medium, $p$ is the Floquet mode index, $k=2\pi/\lambda$, $ k^{(m)}=k\sqrt{\varepsilon^{(m)}}$,  $\theta _{\text{inc}}=0$ is the angle of incidence, $E_0=1$ is the amplitude of the incident plane wave,  and $B_p^{(M)}=0$ to satisfy the radiation condition, where $M$ is the index of the lowest (semi-infinite) medium. The $A_p^{(m)}$ and $B_p^{(m)}$ are unknown coefficients. When $d$ is sufficiently small, $k_{xp}$ will be large, so that $k_{yp}^{(m)}$ will be imaginary when $p \ne 0$, and all terms in the sum of \eqref{eqAppend1} will evanesce except the $p=0$ term. In that case, the amplitude of the sole propagating transmitted wave will be $T=A_0^{(M)}$, which is to be determined.

Solutions of the $A_p^{(m)}$ and $B_p^{(m)}$ unknowns are obtained by solving a system of relations between these unknowns along the interfaces. The components of the electric and magnetic fields tangent to the interfaces between the dielectrics must be continuous across these surfaces.  For the assumed TE polarization, the E-field as given in \eqref{eqAppend1} is directed in the $z$-direction, which is parallel to these interfaces.  Whereas the H-field  possesses both $x$- and $y$-components, only the $x$-component is parallel to these interfaces.  This component is 
\begin{equation}
{H_x}^{(m)}=-\frac{i}{k\eta }\frac{\partial {E_z}^{(m)}}{\partial y},
\label{eqAppend6}
\end{equation}
where $\eta$ is the impedance of free space. The boundary conditions at dielectric interfaces may then be written
\begin{equation}
E_z^{(m)}(x,y^{(m)})=E_{z}^{(m-1)}(x,y^{(m)}),1\le m\le M,
\label{eqAppend7}
\end{equation}
\begin{equation}
H_x^{(m)}(x,y^{(m)})=H_x^{(m-1)}(x,y^{(m)}),1\le m\le M.
\label{eqAppend8}
\end{equation}
Using \eqref{eqAppend3} and \eqref{eqAppend4}, for each Floquet mode $p$ these lead to 
\begin{equation}
A_p^{(m)}+B_p^{(m)}-B_p^{(m-1)}=\delta_{p0},m=1,
\label{eqAppend8a}
\end{equation}
\begin{multline}
A_p^{(m)}+B_p^{(m)}-A_p^{(m-1)}e^{ik_{yp}^{(m-1)}h^{(m-1)}} \\ -B_p^{(m-1)}e^{-ik_{yp}^{(m-1)}h^{(m-1)}}=0,2\le m\le M,
\label{eqAppend8b}
\end{multline}
\begin{multline}
-k_{yp}^{(m)}A_p^{(m)}+k_{yp}^{(m)}B_p^{(m)}-k_{yp}^{(m-1)}B_p^{(m-1)}\\=-k_{yp}^{(m-1)}\delta_{p0},m=1,
\label{eqAppend8c}
\end{multline}
\begin{multline}
-k_{yp}^{(m)}A_p^{(m)}+k_{yp}^{(m)}B_p^{(m)}+k_{yp}^{(m-1)}A_p^{(m-1)}e^{ik_{yp}^{(m-1)}h^{(m-1)}} \\ -k_{yp}^{(m-1)}B_p^{(m-1)}e^{-ik_{yp}^{(m-1)}h^{(m-1)}}=0,2\le m\le M,
\label{eqAppend8d}
\end{multline}

Equations \eqref{eqAppend8a} to \eqref{eqAppend8d} are the boundary conditions that are satisfied at dielectric interfaces. The conditions which must be satisfied across the interface that contains the wire array ($y=y^{(3)}$ in Fig. \ref{figAppend}) will now be discussed. Referring to this interface as $y=y^{(m)}$, it is bounded above by air region $m-1$, and below by air region $m$, so that $\varepsilon^{(m-1)}=\varepsilon^{(m)}=1$, $k^{(m-1)}=k^{(m)}=k$, $k_{yp}^{(m-1)}=k_{yp}^{(m)}=\sqrt{k^{2}-k_{xp}^2}\equiv k_{yp}$.

A current $I$ in each wire is induced by the fields \eqref{eqAppend1} in layers $m-1$ and $m$. Since the field is the sum of Floquet mode components, each such component $p$ will induce a current $I_p$ in each wire. In turn, the field produced in an adjacent air region by the array of wires carrying a current $I_p$ is
\begin{equation}
E_{zp}^{wire}=I_{p}F(x,y),
\label{eqAppend9}
\end{equation}
where
\begin{multline}
F(x,y)=-\frac{k\eta}{4}\sum\limits_{\nu=-\infty }^{\infty }{H_{0}^{(1)}\left[ k\sqrt{{(x-\nu d)^2}+{(y-y^{(m)})^2}}   \right]} \\ =-\frac{k\eta }{2d}\sum\limits_{q=-\infty }^{\infty }{{e^{ik_{xq}x}}}\frac{{e^{ik_{yq}|y-{y^{(m)}}|}}}{k_{yq}},
\label{eqAppend10}
\end{multline}
and $H_0^{(1)}(x)$ is the zeroth order Hankel function of the first kind with agrument $x$. The last sum in \eqref{eqAppend10} is derived from Poisson's formula, and converges much more rapidly than the sum over Hankel functions. 

In \eqref{eqAppend3}, the term containing $A_p^{(m)}$ can be considered a "downward" wave, while that containing $B_p^{(m)}$ an "upward" wave. Since the currents in the wires at $y=y^{(m)}$ are induced by waves incident upon them, these incident waves from layers $m-1$ and $m$ are those with coefficients $A_p^{(m-1)}$ and $B_p^{(m)}$, respectively. Along the surface $y=y^{(m)}$ of the wires, the total field $E_{zp}^{tot}$ for Floquet mode $p$ will be the sum of the incident waves and the scattered waves. To utilize this total field in Ohm's law, it is necessary to evaluate it on the surface of the wire, the effective radius of which is $r_{\text{eff}}=w/4$:
\begin{multline}
E_{zp}^{tot}=e^{ik_{xp}r_{\text{eff}}}[A_p^{(m-1)}e^{-ik_{yp}(-h^{(m-1)})}+B_p^{(m)}]\\+I_pF(r_{\text{eff}},y^{(m)})=I_p\tilde{Z}.
\label{eqAppend11}
\end{multline}
The last equality follows from Ohm's law, where $\tilde{Z}$ is the distributed impedance of the wire that is assumed known. This equation can be solved for $I_p$, and the total current over all Floquet modes would then be $I=\sum\limits_pI_p$:
\begin{equation}
I=\frac{\sum\limits_{p=-\infty }^{\infty }e^{ik_{xp}r_{\text{eff}}}[A_p^{(m-1)}+B_p^{(m)}]}{\tilde{Z}-F},
\label{eqAppend12}
\end{equation}
where $F=F(r_{\text{eff}},y^{(m)})$, and $h^{(m-1)}\rightarrow 0$ has been implemented in anticipation of the air surroundings disappearing later on. The numerical implementation of \eqref{eqAppend10} in \eqref{eqAppend12} requires special handling since for $\nu=0$, $y=y^{(m)}$, $x=r^{\text{eff}}$, the argument of $H_0^{(1)}$ is very small \cite{EpsteinPhysRevAppl2017,RabinovichIEEETAP2018,RabinovichIEEETAP2020}.

The boundary conditions for the field along dielectric interfaces were given in \eqref{eqAppend7} and \eqref{eqAppend8}. For interfaces coinciding with wire arrays, the E-field boundary condition \eqref{eqAppend7} still holds as implemented in \eqref{eqAppend8b} for Floquet mode $p$:
\begin{equation}
A_p^{(m)}+B_p^{(m)}-A_p^{(m-1)}-B_p^{(m-1)}=0,
\label{eqAppend12a}
\end{equation}
where $h^{(m-1)}\rightarrow 0$ has again been implemented. However, unlike \eqref{eqAppend8}, the H-field boundary condition for a wire array expresses the fact that the difference in $H_x$ across the boundary equals the surface current density along it:
\begin{equation}
H_x^{(m)}(x,y^{(m)})-H_x^{(m-1)}(x,y^{(m)})=\frac{I}{d}.
\label{eqAppend13}
\end{equation}
 Invoking \eqref{eqAppend6}, \eqref{eqAppend1}, \eqref{eqAppend3} and \eqref{eqAppend12}, this may be written for Floquet mode $p$,
\begin{multline}
k_{yp}[-A_p^{(m)}+B_p^{(m)}+A_p^{(m-1)}-B_p^{(m-1)}] \\=\sum\limits_{q=-\infty }^{\infty }\alpha_q[A_q^{(m-1)}+B_q^{(m)}],
\label{eqAppend14}
\end{multline}
where $\alpha_q=\frac{k\eta}{d}\frac{e^{ik_{xq}r_{\text{eff}}}}{\tilde{Z}-F}$. 

Equations \eqref{eqAppend8a} to \eqref{eqAppend8d}, \eqref{eqAppend12a} and \eqref{eqAppend14} represent a system of linear equations in the unknowns $A_p^{(m)}$, $B_p^{(m)}$. In order to solve them, the infinite sums must be truncated: $\sum\limits_{p=-\infty }^{\infty }\rightarrow \sum\limits_{p=-P}^{P}$, implying a total of $2P+1$ Floquet modes. Consider the six-layer structure of Fig. \ref{figAppend}. In the incident $m=0$ region there are $2P+1$ unknowns $B_p^{(0)}$, and in the transmission $m=M=5$ region there are $2P+1$ unknowns $A_p^{(M)}$. In the remaining regions $1\le m\le M-1$, there are $2(2P+1)$ in each:  $A_p^{(m)}$ and $B_p^{(m)}$. There are therefore a total of $2M(2P+1)$ unknowns. The equations containing these unknowns have been derived from the interfaces between the regions. There are $M$ such interfaces at $y=y^{(m)},1\le m\le M$. At each interface, there are two conditions to be satisfied, one related to the E-field, the other related to the H-field. Each of these equations can be written for $2P+1$ values of $p$, so there are the same number of equations as unknowns.  The problem can therefore be solved uniquely for the $A_p^{(m)}$ and $B_p^{(m)}$, and in particular for the transmission coefficient $T=A_0^{(M)}$.

\section*{Acknowledgment}

The authors are greatly indebted to Doron Rainish, Heylal Mashaal, Roni Ashkenazi, Dotan Goberman and Avi Freedman for their "brainstorming" contibutions, CST calculations, and suggestions.

\bibliographystyle{IEEEtran}
\bibliography{MDARC} 

\end{document}